\documentclass[journal=jacsat, manuscript=article, layout=twocolumn]{achemso}

\usepackage[fontsize=10pt]{fontsize}
\usepackage[utf8]{inputenc}
\usepackage[T1]{fontenc}
\usepackage{amsmath, amssymb}
\usepackage{graphicx}
\usepackage{siunitx}
\usepackage{xcolor}
\usepackage{float}
\usepackage{tikz}
\usepackage{hyperref}
\DeclareSIUnit\wn{\cm\tothe{-1}}
\DeclareSIUnit\calorie{cal}

\usepackage{tabularx}
\newcolumntype{C}[1]{>{\centering\arraybackslash\hspace{0pt}}p{#1}}
\newcolumntype{Y}{>{\raggedright\arraybackslash}X}

\newcommand{\geom}{adsorption environment}
\newcommand{\geometries}{adsorption environments}

\newcommand{\review}[1]{\textcolor{black}{#1}}

\let\oldnameref\nameref
\renewcommand{\nameref}[1]{\textit{\oldnameref{#1}}}

\let\oldmaketitle\maketitle
\let\maketitle\relax

\emergencystretch=\maxdimen
\title{Vibrational Energy Dissipation in Non-Contact Single-Molecule Junctions Governed by Local Geometry and Electronic Structure}

\author{Lukas H\"ormann}
\email{lukas.hoermann@warwick.ac.uk}
\affiliation{Department of Chemistry, University of Warwick, Gibbet Hill Rd, Coventry, CV4 7AL, United Kingdom}
\altaffiliation{Department of Physics, University of Warwick, Gibbet Hill Rd, Coventry, CV4 7AL, United Kingdom}

\author{Reinhard J. Maurer}
\email{r.maurer@warwick.ac.uk}
\affiliation{Department of Chemistry, University of Warwick, Gibbet Hill Rd, Coventry, CV4 7AL, United Kingdom}
\altaffiliation{Department of Physics, University of Warwick, Gibbet Hill Rd, Coventry, CV4 7AL, United Kingdom}

\begin{tocentry} 
\includegraphics{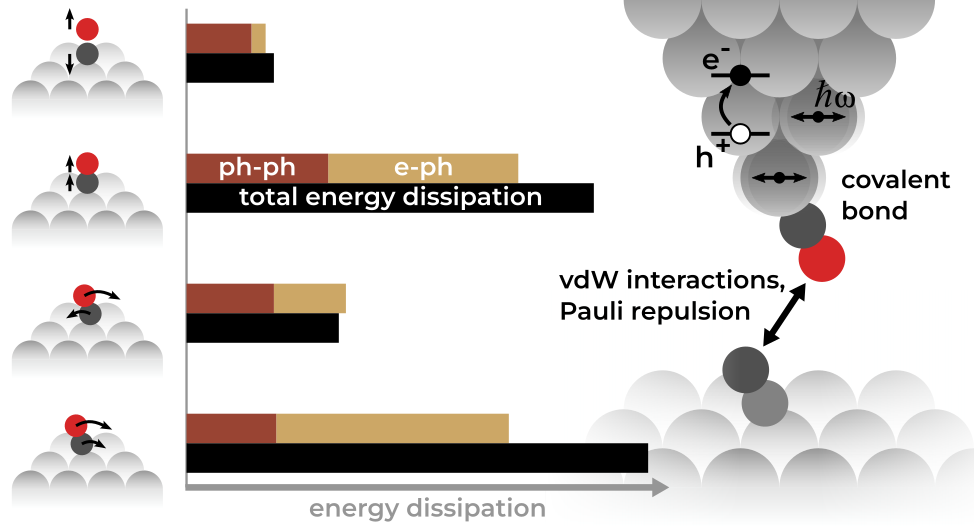}
\end{tocentry}

\begin{document}

\twocolumn[
\begin{@twocolumnfalse}
\oldmaketitle
\begin{abstract}
The vibrational dynamics of adsorbate molecules in single-molecule junctions depend critically on the geometric structure and electronic interactions between molecule and substrate. Vibrations, excited mechanochemically or by external stimuli, dissipate energy into substrate electrons and phonons. Energy dissipation leads to the broadening of spectral lines, vibrational lifetimes, and the coupling between molecular and substrate phonons. It affects molecular manipulation, giving rise to nanoscale friction, and contributes to scanning probe and surface spectroscopy signals. We present an approach to disentangle adsorbate vibrational dynamics in non-contact junctions by employing Density Functional Theory, machine learning, and non-adiabatic molecular dynamics. Focusing on the CO-functionalised Cu surfaces representing a single-molecule junction, a widely studied system in scanning probe and energy dissipation experiments, we reveal strong vibrational mode specificity governed by the interplay of electron-phonon and phonon-phonon coupling. Electron-phonon relaxation rates vary by two orders of magnitude between modes and sensitively depend on the tip-substrate geometry. We find evidence of a weak non-additive effect between both energy dissipation channels, where electron-phonon coupling enhances phonon-phonon coupling. Our predicted vibrational lifetimes agree with infrared spectroscopy and helium scattering experiments. Finally, we outline how our findings can inform and enhance spectroscopy and scanning probe experiments.
\end{abstract}


\vspace{0.3cm}
\end{@twocolumnfalse}
]

\section{Introduction}

The vibrational dynamics of atoms and molecules adsorbed on a surface provide valuable insights into the interactions and energy exchange between adsorbate and substrate. These dynamics are governed by interactions between adsorbates and the surface and quantum-mechanical energy dissipation mechanisms.\cite{tully1993electronic, park2014fundamental} Energy dissipation mechanisms in non-contact single-molecule junctions significantly affect measured signals in atomic force microscopy (AFM), lateral force microscopy (LFM),\cite{okabayashi2023dynamic, nam2024exploring} scanning tunnelling microscopy (STM),\cite{stipe1998single} inelastic electron tunnelling spectroscopy (IETS), and infrared spectroscopy experiments.\cite{hirschmugl1994low, hirschmugl1995chemical} In such experiments, a key challenge is to disentangle geometric structure, dynamics, and energy dissipation mechanisms in the measured signal: For example, AFM-induced hopping of atoms and molecules leads to measurable energy dissipation,\cite{lauhon1999single, komeda2002lateral, nakamura2003hopping, okabayashi2023dynamic} yet the underlying dissipation pathways remain unclear. Helium atom scattering (HAS) experiments of CO adsorbed on Cu surfaces have found a strong dependence of vibrational frequencies on both the surface facet and the presence of defects.\cite{braun1996he}

It is well-established that the surface geometry plays a critical role in shaping molecule-surface interactions and energy transfer processes: Previous studies have shown the effect of defects on static interfacial charge transfer,\cite{wruss2019impact} while adsorption site and facet affect the vibrational properties of adsorbates.\cite{yudanov2003co} Establishing an understanding of how surface geometry alters different energy dissipation channels and affects the vibrational dynamics of molecules adsorbed on surfaces remains an important research area.\cite{sader2012spring}

Figure \ref{fig:experimentalsetup} shows a schematic setup of an idealised single-molecule junction as it would be present in an AFM experiment. If vibrations of the molecule in the junction are excited, for example, through the motion of the tip, energy dissipation away from the molecule can occur by excitation of phonons or electron-hole-pair (EHP) excitations in the metal tip or substrate. The former is due to anharmonicity that induces phonon-phonon coupling (PPC) between adsorbate and substrate.\cite{park2014fundamental} The latter is due to coupling between electronic and nuclear motions, commonly referred to as electron-phonon coupling (EPC).\cite{hirschmugl1994low} Intramolecular vibrational relaxation and radiative processes, such as infrared emission, also contribute to energy dissipation, though their role is typically minor for small adsorbates.\cite{walton1977triboluminescence, dai1995laser} However, for larger adsorbates, intramolecular vibrational relaxation can become a significant dissipation channel.\cite{ge2018electron} Consequently, it is of great interest to explore how the geometric structure of a surface or junction affects the mechanical and electronic interactions with the adsorbate, which govern the relative contributions of EPC and PPC to the vibrational dynamics. Gaining control over these energy dissipation channels requires the ability to identify their distinct signatures and disentangle their interplay---an essential step towards rational design of surface functionality at the atomic scale.

In this work, we study competing energy dissipation mechanisms and their effect on the vibrational dynamics of adsorbate molecules for the exemplary case of CO-decorated tip-surface copper junctions. CO on Cu is a prototypical surface-adsorbate system. It serves as a benchmark in surface science, enabling the systematic investigation of fundamental processes such as adsorption, diffusion, and vibrational energy dissipation.\cite{hollins1979interactions, hirschmugl1994low, weymouth2014quantifying, gameel2018unveiling, okabayashi2023dynamic} Moreover, it is widely used in atomic-resolution scanning probe measurements.\cite{gross2009chemical} Owing to the similar electronic structure of coinage metals, insights from CO on Cu are transferable to Ag and Au surfaces, making it relevant for experiments and applications in catalysis and energy conversion, where electron–phonon and phonon–phonon coupling govern surface reactivity.\cite{HAMMER1995211} Various previous theoretical studies of energy dissipation for CO on different, idealised Cu surfaces\cite{maurer2016ab, lorente2005co, persson2004theory} have been reported, but no clear analysis of the role of the local adsorption geometry exists to date. A specific challenge in this system for electronic structure theory is the correct description of the potential energy surface of CO on Cu(111).\cite{gajdovs2004co, gajdovs2005co, feibelman2001co} We address this by using the state-of-the-art screened hybrid functional HSE06\cite{heyd2003hybrid} in combination with the MBD-NL van der Waals (vdW) correction\cite{hermann2020density}. From a first-principles perspective, we describe the details and main energy dissipation mechanisms, namely EPC and PPC, which govern the dynamics at the atomic scale.\cite{park2014fundamental} We find that the surface geometry plays a significant role in these mechanisms, with flatter surfaces exhibiting stronger EPC and PPC. Additionally, we observe a strong mode dependence of EPC, with relaxation rates differing by two orders of magnitude. The molecular dynamics with electronic friction (MDEF) approach\cite{tully1990molecular, maurer2016ab} allows us to study the interplay of the two dissipation channels and their concerted effects on the relaxation dynamics. We find signs of weak non-additivity between the two effects, suggesting that subtle dynamical steering may arise from one or both mechanisms. Our results are in good agreement with existing experiments and support the interpretation of single-molecule dynamics in non-contact junctions.

\section{Results}

\subsection{Models of the tip geometry in non-contact junctions}
\label{sec:modelling_the_dynamics}

\begin{figure}
	\centering
	\includegraphics[width=\linewidth]{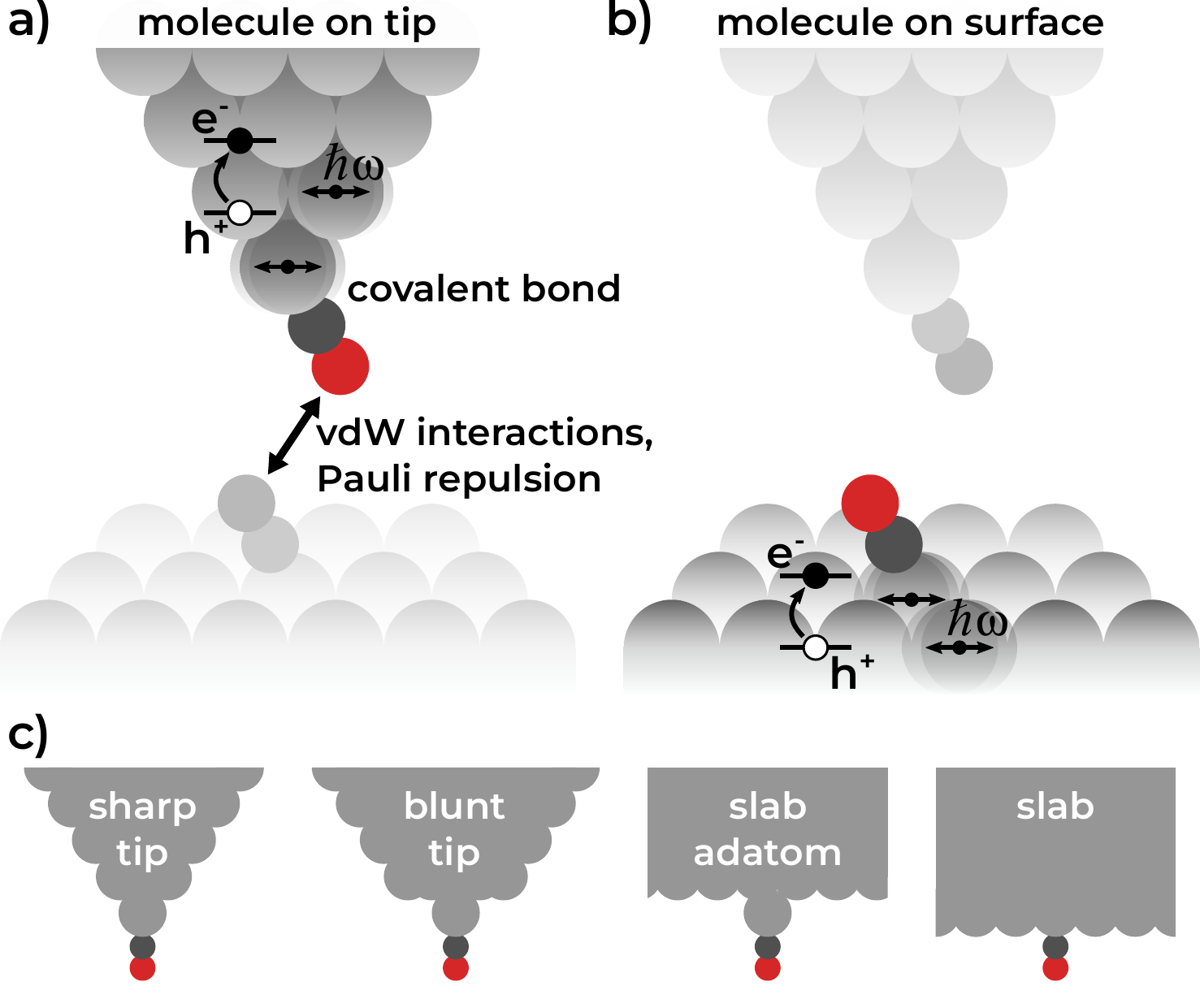}
	\caption{Schematic non-contact single-molecule junction experiment where a CO molecule on a surface is probed with a CO-decorated tip; Van-der-Waals (VdW)-interactions and Pauli-repulsion govern the interaction between sample and probe; Covalent bonds are present between the CO molecule and the metallic component; Energy exchange occurs primarily between chemically bonded components, in the form of EPC and PPC; a) Molecule is attached to the tip; b) Molecule is attached to the surface; c) Different {\geometries} considered in this work; All systems are calculated with periodic boundary conditions, with the depicted tip geometries extended by a further four layers of metal slab (not shown here).}
	\label{fig:experimentalsetup}
\end{figure}

As shown in Figure \ref{fig:experimentalsetup}, we can consider two interaction regimes for a non-contact CO tip-molecule-surface junction: In one case, the molecule is attached to the surface, in the other case, it is attached to the tip. The molecule can be excited by a variety of processes: The motion of the tip will displace the molecule with respect to the corrugated potential energy landscape of the underlying substrate.\cite{hapala2014origin} The mechanochemical forces the molecule experiences in the process will excite its vibrations. The molecular vibrations can absorb light,\cite{hirschmugl1994low} supported by field enhancement in the junction.\cite{peller2021quantitative} A potential bias and electron tunnelling can lead to current-induced forces.\cite{okabayashi2023dynamic} In all cases, molecular vibrations will be excited, leading to energy dissipation into EHPs and phonons in the metal. The case where the molecule is attached to the surface is not too different from the scenario in the absence of the tip. The tip and molecule are separated by a distance of multiple Angstroms, so mechanochemical forces on the molecule will be weak. Other perturbation scenarios have previously been studied.\cite{persson2004theory, lorente2005co} For simplicity, we consider only surfaces in both the slab and slab–adatom systems that correspond to the Cu(111) facet.

We turn our attention to the second case, where the molecule is attached to the tip. Regardless of the origin of the vibrational excitation, energy dissipation into the underlying substrate in this scenario will be almost negligible, as evidenced by calculations for CO on Cu in Section S1 of the Supporting Information. Therefore, molecular vibrations will dominantly dissipate energy into EHPs and phonons in the tip. This motivates our model design: We consider four different possible {\geometries}: CO on a sharp tip (sharp tip), CO on a blunt tip (blunt tip), CO on an adatom on a surface slab (slab adatom), and CO on a slab (slab), which are depicted in Figure \ref{fig:experimentalsetup}c and Figure S3 in the Supporting Information. The underlying surface, in the case of lateral motion of the tip, only serves as a corrugated energy landscape that triggers molecular vibrations\cite{nam2024exploring} and therefore does not have to be explicitly considered in the atomistic model. The ``tip'' models are therefore represented by two-dimensional periodic surface slab geometries. Further details on the models and the computational methodology are provided in the \textit{Methods} and Section S1 of the Supporting Information.

\subsection{Vibrational mode analysis}

\begin{figure}[H]
	\centering
	\includegraphics[width=\linewidth]{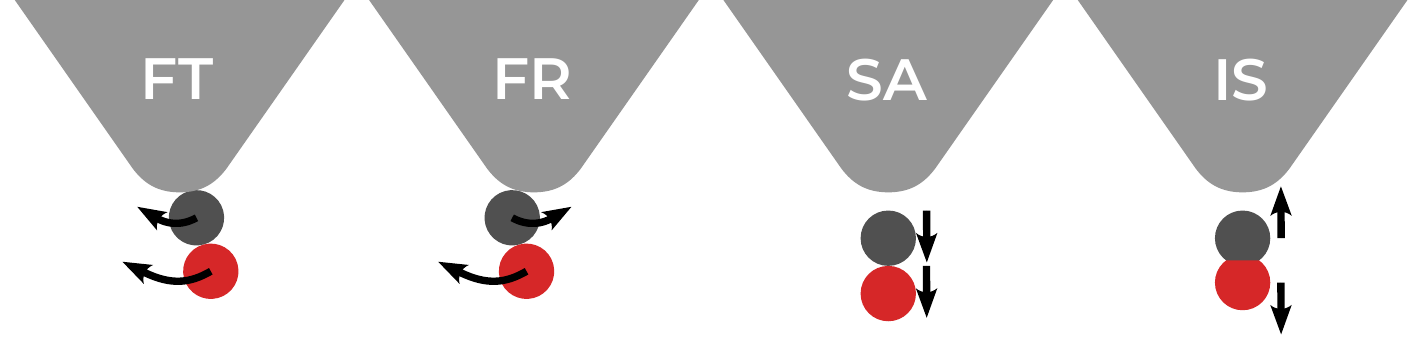}
	\caption{Four eigenmodes of CO on Cu surface; Frustrated rotation (FR); Frustrated translation (FT); Surface adsorbate stretch (SA); Internal stretch of CO bond (IS).}
	\label{fig:vibration_modes}
\end{figure}

CO adsorbed on a flat metal surface has four distinct types of vibrational modes: frustrated translation (FT), frustrated rotation (FR), surface-adsorbate stretch (SA), and the internal stretch of the CO bond (IS), as illustrated in Figure \ref{fig:vibration_modes}. The FT and FR modes consist of two degenerate modes each.

We determine vibrational energies from second-order force constants (i.e. the Hessian) for all atoms in the molecule and the top layers of the four surface models. Simulations include ten layers of freely movable Cu atoms, spanning hundreds of degrees of freedom. Energies, forces, and second-order force constants are evaluated with a MACE graph neural network model\cite{batatia2022mace} trained on first-principles data (see \textit{Methods}). The MACE potential accurately reproduces the underlying HSE06 data, correctly predicting the adsorption sites: the top site for CO on a (111) surface slab of copper and the hollow site for the adatom in the CO-adatom arrangement (see Section S2 of the Supporting Information). The calculated vibrational frequencies for CO on the Cu(111) slab show good agreement with experimental data, as presented in Table \ref{tab:vibration_frequencies}. Interestingly, the stiffness of low-energy modes increases as the tip becomes blunter, as indicated in Table \ref{tab:vibration_frequencies}. Moreover, there is a jump in stiffness when transitioning from CO adsorbed on an adatom to CO adsorbed directly on the surface. HAS experiments estimated the energy of the FT modes on a flat Cu(111) surface to be \SI{4.01}{\milli\electronvolt}, while the energy dropped to \SI{3.00}{\milli\electronvolt} at step edges of the Cu(211) and Cu(511) surfaces.\cite{braun1996he} This trend is in good agreement with our finding that more tapered (or corrugated) {\geometries} exhibit lower vibrational energies for the FT mode. 

\begin{table}
    \setlength\tabcolsep{0.0cm}
	\centering
	\caption{Vibrational energies of CO molecule in different adsorption geometries; Experimental vibrational energies of the CO molecule on a Cu(111) surface.~\cite{braun1996he, hirschmugl1995chemical}}
	\begin{tabularx}{\linewidth}{Y|C{1.3cm}|C{1.3cm}|C{1.3cm}|C{1.3cm}}
		& \multicolumn{4}{c}{\textbf{vibrational energies / meV}} \\
		& \textbf{FT} & \textbf{FR} & \textbf{SA} & \textbf{IS} \\\hline
        sharp tip & 2.74 & 31.38 & 49.06 & 271.48 \\
		blunt tip & 2.89 & 32.28 & 50.35 & 270.86 \\
		slab adatom & 3.17 & 31.74 & 49.13 & 272.26 \\
		slab top & 5.29 & 35.92 & 47.97 & 264.51 \\\hline
		slab top exp.\cite{braun1996he, hirschmugl1995chemical} & 4.07 & 36.47 & 41.41 & 257.63 \\
	\end{tabularx}
	\label{tab:vibration_frequencies}
\end{table}

\begin{figure}
	\centering
	\includegraphics[width=\linewidth]{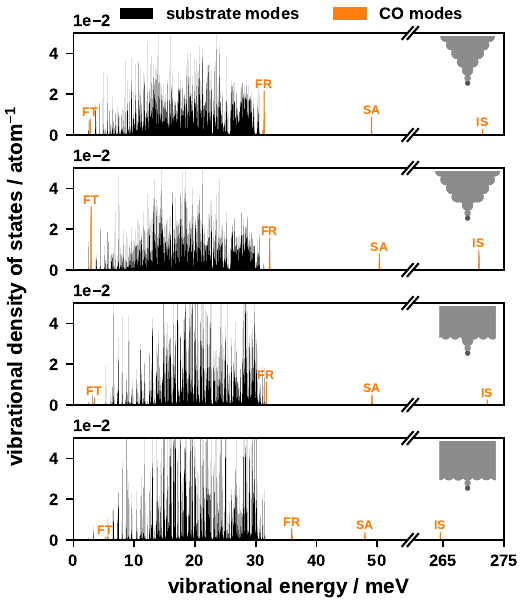}
	\caption{VDOS of the four different {\geometries}; The total VDOS is shown in black; The contribution of the CO-modes is shown in orange.}
	\label{fig:vdos}
\end{figure}

The vibrational density of states (VDOS) (see Figure \ref{fig:vdos}) provides an overview of all vibrational modes. The VDOS is determined from the same molecular dynamics (MD) simulations that yield the results presented in \nameref{sec:phonon_phonon_coupling}. We differentiate between modes localised dominantly at the molecule (orange in Figure \ref{fig:vdos}), where the majority of the kinetic energy is contained in the CO and substrate modes (black in Figure \ref{fig:vdos}), where most of the energy is in the motion of the Cu atoms. The FR, SA, and IS vibrational modes of CO exhibit higher vibrational energy compared to most copper substrate modes, which typically range between approximately \SI{5}{\milli\electronvolt} and \SI{30}{\milli\electronvolt}. We note that the VDOS shown here is normalised by the number of atoms. The substrate vibrational states are more localised in energy (or frequency) for slab and slab-adatom surfaces and more delocalised for the sharp and blunt tips. For sharp and blunt tips, a significant portion of the surrounding space is vacuum, whereas this space is occupied by atoms in the slab and slab-adatom surfaces. The presence of more atoms in these flatter {\geometries} results in a greater number of substrate vibrational states per volume. This facilitates stronger overall coupling between the CO and substrate vibrational modes, provided that the CO and Cu degrees of freedom are sufficiently coupled. As a rule of thumb, if the off-diagonal elements of the Hessian (for CO-Cu coupling) are of a similar magnitude, we can expect strong coupling. As shown in Figure S4 in the Supporting Information, this is indeed the case. In the case of the sharp and blunt tip, we find a strong coupling for the CO to a small number of Cu degrees of freedom. The slab adatom and slab geometries exhibit weaker coupling but a larger number of states. When discussing PPC, we will return to this observation.

\subsection{Phonon-phonon coupling}
\label{sec:phonon_phonon_coupling}

The atomic motion of the CO molecule is coupled to the atomic motion of the Cu substrate through phonon-phonon interactions and to EHP through EPC.\cite{park2014fundamental} Even at low temperatures, PPC plays a significant role in the dynamics of molecular vibrations. In our study, we use a temperature of \SI{5}{\kelvin}, which is commonly employed in experiments such as AFM measurements with CO-decorated tips. PPC is determined by analysing the dynamics of a CO molecule on {\geometries} where approximately 500 Cu atoms are free to move per unit cell, allowing energy transfer between CO modes and substrate modes. To determine coupling strengths and vibrational lifetimes, we use vibrational analysis of MD trajectories in thermodynamic equilibrium.\cite{sun2014dynamic} We refer to this approach as \textit{equilibrium correlation analysis}. A conceptually related approach, which combines vibrational projection, statistical analysis, and local quantum vibrational embedding, was recently introduced to investigate ultrafast energy transfer pathways from lattice phonons to intramolecular modes in molecular crystals.\cite{doi:10.1021/jacsau.4c00775} We first thermalise each system at a temperature of \SI{5}{\kelvin} and then allow it to evolve in thermodynamic equilibrium over time. Over time, the velocity time series of the atoms becomes decorrelated, as the atoms transfer mechanical energy between each other. The rate of this decorrelation can be found in the cross-correlation function. For this analysis, we employ normal mode decomposition to project the velocity time series onto the basis defined by the vibrational eigenvectors. For the velocity time series of two modes $v_\mathrm{i}(t)$ and $v_\mathrm{j}(t)$ the velocity cross-correlation function $C_\mathrm{ij}(\tau)$ is as follows:
\begin{equation}
	C_\mathrm{ij}(\tau) = (v_\mathrm{i} * v_\mathrm{j})(\tau) = \int_{-\infty}^{\infty} v_\mathrm{i}^*(t) v_\mathrm{j}(t-\tau) dt
\end{equation}

The Fourier transform of the velocity cross-correlation function yields the cross-spectrum:
\begin{equation}
    \tilde{C}_\mathrm{ij}(\omega) = \int_{-\infty}^{\infty} C_\mathrm{ij}(\tau) e^{-i \omega \tau} d\tau
\end{equation}

The cross-spectrum provides valuable insights into vibrational lifetimes and relative coupling strengths. Let us first discuss the coupling strength, which can be determined by analysing the off-diagonal elements $\tilde{C}_\mathrm{ij}(\omega)$. These exhibit two Lorentzian-shaped peaks at the characteristic frequencies of modes $i$ and $j$, with their heights reflecting the coupling strength between these modes (see Section S6.2 of the Supporting Information). To further understand this interaction, we examine the coupling density $\rho_\mathrm{j}(\omega)$, defined as the sum of the cross spectra $\tilde{C}_\mathrm{ij}(\omega)$ for all modes $j$ correlated with the mode of interest $i$:
\begin{equation}
	\rho_\mathrm{i}(\omega) = \sum_\mathrm{j} \tilde{C}_\mathrm{ij}(\omega)
\end{equation}

\begin{figure}
	\centering
    \includegraphics[width=\linewidth]{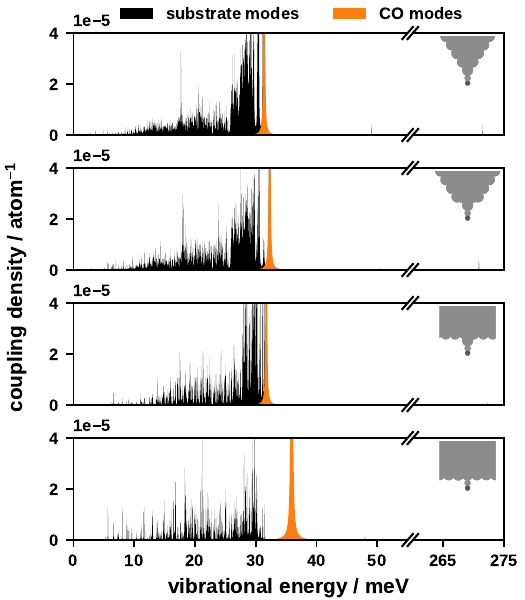}
	\caption{Coupling strength of the FR mode to other vibrational modes as a result of PPC.}
	\label{fig:coupling_dos_FT}
\end{figure}

A broad peak in the coupling density at a given frequency means that modes at this frequency are correlated with the mode of interest, $\omega_\mathrm{i}$, indicating strong vibrational coupling. Our analysis reveals that the coupling is strongest with adjacent substrate vibrational modes, as shown in Figure \ref{fig:coupling_dos_FT} and Figures S24, S25, S26, and S27 in the Supporting Information. This effect is less pronounced in case of the energetically decoupled SA and IS modes. The coupling between different CO modes is small in comparison, indicating that intramolecular vibrational redistribution plays only a minor role. Notably, the FR and IS modes exhibit the strongest coupling.

The geometry of the surface has a substantial impact on the coupling density. Among the different geometries, the sharp tip exhibits the weakest coupling, while the slab geometry has the strongest coupling density. This observation can be attributed to the VDOS discussed earlier. As surface geometry becomes more atomically flat, the VDOS increases, offering more vibrational states to accommodate the energy of CO modes and thereby enhancing PPC.

The diagonal terms $\tilde{C}_\mathrm{ii}(\omega)$ of the cross-correlation function contain the power spectrum. The power spectrum exhibits a single Lorentzian-shaped peak at the characteristic frequency of the mode $i$, where the full width at half maximum $\Delta f_\mathrm{i}$ is inversely proportional to the lifetime of mode $\tau_\mathrm{i}$.
\begin{equation}
    \tau_\mathrm{i} = \frac{1}{\Delta f_\mathrm{i}}
\end{equation}

As shown in Table \ref{tab:lifetimes_phph}, the overall lifetimes tend to decrease as the surface becomes more atomically flat. Interestingly, this trend does not apply uniformly to all modes. The FT mode shows the strongest dependence on surface geometry, while this dependence is less pronounced for the FR and SA modes. The lifetime of the IS mode remains largely unaffected by changes in surface shape. Overall, the influence of the surface geometry is strongest in low-energy modes and becomes weaker as the energy of the mode increases. The largest phononic lifetime (FT on sharp tip model) is one order of magnitude larger than the smallest lifetime (FT on slab model).

\begin{table}
    \setlength\tabcolsep{0.0cm}
	\centering
	\caption{CO adsorbate vibrational lifetimes due to PPC.}
	\label{tab:lifetimes_phph}
	\begin{tabularx}{\linewidth}{Y|C{1.3cm}|C{1.3cm}|C{1.3cm}|C{1.3cm}}
		& \multicolumn{4}{c}{\textbf{Lifetime / ps}} \\
		& \textbf{FT} & \textbf{FR} & \textbf{SA} & \textbf{IS} \\\hline
		sharp tip & 142.1 & 69.0 & 56.2 & 48.5 \\
		blunt tip & 110.9 & 67.0 & 31.9 & 65.4 \\ 
		slab adatom & 59.8 & 65.3 & 53.7 & 46.3 \\ 
		slab & 19.3 & 24.6 & 26.8 & 53.7 \\
	\end{tabularx}
\end{table}

\subsection{Electron-phonon coupling}
\label{sec:electron_phonon_coupling}

EPC is a significant factor in vibrational energy dissipation on metal surfaces.\cite{ARNOLDS201045, park2014fundamental} Due to the absence of a band gap, the Born-Oppenheimer approximation does not strictly hold, and the motion of adsorbate atoms can induce EHP excitations that lead to nonadiabatic effects, sometimes also called electronic friction effects. In the weak perturbation regime, e.g. where adsorbate motion is triggered mechanochemically or through vibrational excitation, this effect leads to nonadiabatic energy dissipation. 
To describe the impact of these excitations on dynamics, different simulation methods have been developed. Time-dependent perturbation theory has emerged as a feasible approach for extended systems. This method enables the calculation of the electron-phonon-induced vibrational lifetimes (see \nameref{sec:description_of_electron_phonon_coupling} in the \textit{Methods}),\cite{box2023ab} which we discuss in the following.

\begin{figure}[h]
	\centering
	\includegraphics[width=\linewidth]{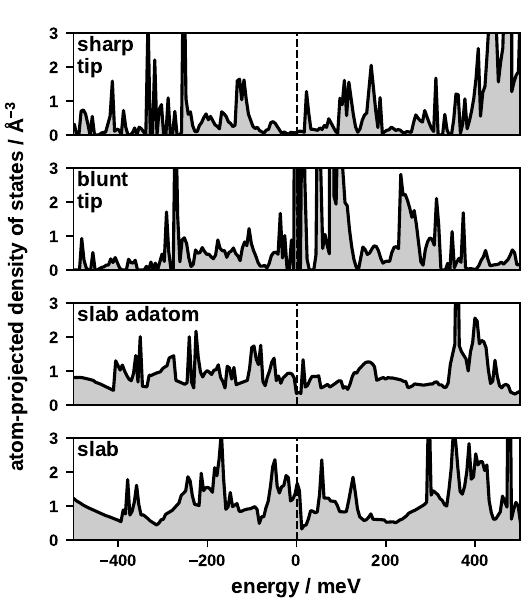}
	\caption{DOS projected at the Cu atom to which the CO molecule bonds; Flatter surface geometry shows a more delocalised projected DOS.}
	\label{fig:dos}
\end{figure}

We evaluate EPC at a temperature of \SI{5}{\kelvin}, typical for non-contact single-molecule junction experiments. At such low temperatures, EPC dominates over PPC in metals \cite{maurer2017mode, box2020determining} and first-order EPC is the dominant contribution to vibrational lifetimes, allowing higher-order EPC effects to be neglected.\cite{PhysRevLett.120.156804} Its strength is sensitive to the electronic states that interact with nuclear motion, which we characterize by the electronic density of states (DOS) projected onto the copper atom bonded to CO (PDOS), which is obtained from Density Functional Theory (DFT). The PDOS is highly sensitive to the geometry of the system. Since the energy to excite electrons is supplied by vibrational states with frequency $\omega$, only a small portion of the DOS near the Fermi level ($E_F \pm \hbar\omega$)---relative to the energy scale of electronic states---is accessible. Therefore, analysing the DOS at the Fermi level is instructive for understanding the strength of EPC. As shown in Figure \ref{fig:dos}, the PDOS of the sharp and blunt tips exhibits states localised in energy. The magnitude of the PDOS close to the Fermi level is small for the sharp tip, while the blunt tip exhibits energetically localised spikes. In contrast, the slab and slab-adatom systems have a PDOS that is delocalized in energy. The slab PDOS exhibits a larger magnitude compared to that of the adatom on the slab. Since a larger magnitude of the PDOS at and close to the Fermi level allows for stronger EPC, we expect that EPC will be stronger at flatter surfaces.

The strength of EPC is reflected in the vibrational lifetimes, where shorter lifetimes indicate stronger coupling (see Table \ref{tab:lifetimes_eph}). The FT and SA modes experience relatively low EPC (resulting in longer lifetimes), whereas FR and IS modes exhibit EPC approximately an order of magnitude higher (with lifetimes an order of magnitude shorter) compared to FT and SA. Previous studies of CO on different Cu facets\cite{persson2004theory, lorente2005co, maurer2016ab} have reported similar trends of mode dependence, as we show in Table \ref{tab:lifetimes_eph}. Our results for the slab are in good agreement with literature values. In the case of the slab, there is an ongoing debate about whether the FR or the IS decays faster. Indeed, the lifetimes reported here and in other literature are very similar for these modes. However, as the tip becomes more corrugated, the IS mode clearly becomes the fastest dissipation mode, while the lifetime of the FR mode increases. As shown in Table \ref{tab:lifetimes_eph}, vibrational lifetimes are significantly longer on sharper tips, with CO on a flat surface exhibiting the shortest lifetimes. This trend aligns with the reduced magnitude of the PDOS close to the Fermi level on more corrugated surfaces. 

\begin{table}[h]
    \setlength\tabcolsep{0.0cm}
	\centering
	\caption{Electron-phonon-induced vibrational lifetimes in the harmonic approximation, including a comparison to theoretical values previously reported in literature.}
	\label{tab:lifetimes_eph}
	\begin{tabularx}{\linewidth}{Y|C{1.3cm}|C{1.3cm}|C{1.3cm}|C{1.3cm}}
		& \multicolumn{4}{c}{\textbf{Lifetime / ps}} \\
		& \textbf{FT} & \textbf{FR} & \textbf{SA} & \textbf{IS} \\\hline
        sharp tip & 185.9 & 21.2 & 80.0 & 2.8 \\ 
		blunt tip & 99.1 & 6.0 & 39.0 & 3.3 \\
		slab adatom & 110.8 & 10.8 & 52.9 & 4.9 \\
		slab (Cu(111))& 47.9 & 3.6 & 12.5 & 1.9 \\\hline
        CO on Cu(111)\cite{persson2004theory} & 84.4 & 2.1 & 7.8 & 2.2 \\
		CO on Cu(110)\cite{lorente2005co} & 76.9 & 2.7 & 10.8 & 2.2 \\
        CO on Cu(100)\cite{maurer2016ab} & 125.0 & 2.6 & 20.0 & 3.2 
	\end{tabularx}
\end{table}

\subsection{Electron-phonon and phonon-phonon coupling}
\label{sec:electron_phonon_and_phonon_phonon_coupling}

\begin{figure}
	\centering
	\includegraphics[width=\linewidth]{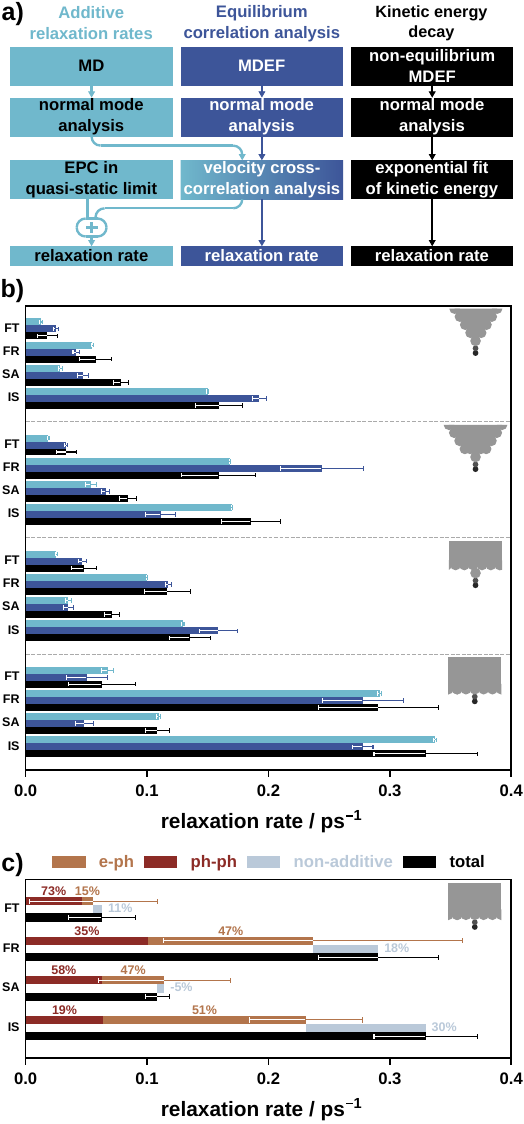}
	\caption{a) Flow charts showing the three different workflows used to calculate relaxation rates in this work; b) Relaxation rates of CO vibrational modes on different \review{{\geometries}} determined with the three methods shown in panel a; c) Contributions to relaxation rates by EPC and PPC when determined using the \textit{kinetic energy decay} method: Light brown: electron-phonon relaxation rate extracted from MDEF; dark-brown: phonon-phonon relaxation rate determined with MD; slate blue: non-additive difference of the sum of ECP and PPC with respect to total relaxation rate; black: total relaxation rate from MDEF.}
	\label{fig:lifetime_methods}
\end{figure}

The experimentally measured lifetime is governed by both EPC and PPC. To capture this combined effect, we employ Langevin dynamics, specifically MDEF.\cite{tully1990molecular} We use a classical approximation, neglecting the zero-point energy of the phonons. This approach has been successfully applied to the dynamics of atoms\cite{askerka2016role, 10.1063/1.4931669} and molecules\cite{maurer2017mode, box2020determining} at metal surfaces. In MDEF, EPC is treated as a friction force represented by the electronic friction tensor (see Equation \ref{eq:langevin_equation}). The electronic friction tensor is calculated in the Markovian quasi-static limit (see \nameref{sec:electron_phonon_coupling}). Here, the frictional force experienced by the nuclei depends only on their instantaneous velocities, and memory effects are ignored. This assumes that the electronic structure adjusts instantaneously to the nuclear configuration. In the quasi-static limit, the electronic friction tensor is computed at zero frequency in the harmonic approximation for a given nuclear geometry without explicitly propagating electronic dynamics. This approach is commonly employed for MDEF on metals.\cite{box2023ab, spiering2019orbital} We determine vibrational relaxation rates and lifetimes using three different methods (see Figure \ref{fig:lifetime_methods}a):

\textbf{Equilibrium correlation analysis.}
Lifetimes are determined by analyzing the velocity cross-correlation function from an MD (without EPC) or MDEF (with EPC) trajectory in thermodynamic equilibrium. This method is introduced and applied in \nameref{sec:phonon_phonon_coupling} to determine PPC relaxation rates and lifetimes from MD trajectories. When EPC is included via MDEF, the method yields total relaxation rates that capture both PPC and EPC.

Table \ref{tab:lifetimes_eph_phph} shows the vibrational lifetimes from \textit{equilibrium correlation analysis} of the MDEF simulation results. While the IS and FR lifetimes are dominated by EPC, the FT mode is dominated by PPC. The two effects are somewhat equal in contribution in the case of the SA mode lifetime. Lifetimes for the FR and IS modes are in good agreement with infrared spectroscopy experiment. However, the experimental lifetime of the SA mode is \SI{2.5}{\pico\second}, while the calculated lifetime is \SI{20.7}{\pico\second} for the Cu(111) surface. A similar discrepancy has been reported in literature and was attributed to inhomogeneous broadening:\cite{hirschmugl1994low} Experiments conducted on rougher surfaces show a greater sensitivity of the mode amplitude and linewidth of the SA mode (and the amplitude of the FR mode) than for the IS mode. The literature lifetimes from infrared spectroscopy experiments should be considered as lower bounds (marked by asterisks (\textsuperscript{*}) in Table \ref{tab:lifetimes_eph_phph}). This is because the observed linewidths are additionally broadened due to limitations of the measurement setup. Since lifetime is inversely related to linewidth, this broadening leads to a systematic underestimation of the lifetimes. Additionally, helium scattering experiments have measured the lifetime of the FT mode of CO on Cu(100) to \SI{8}{\pico\second},\cite{graham1996observation} which is in the same order of magnitude as the lifetime for the FT mode of CO on a Cu(111) slab reported here.

\begin{table}
    \setlength\tabcolsep{0.0cm}
	\centering
	\caption{Vibrational lifetimes due to EPC and PPC calculated by MDEF \textit{equilibrium correlation analysis} compared to experimental lifetimes; Asterisks (\textsuperscript{*}) indicate that experiments correspond to lower bounds on the vibrational lifetime.}
	\label{tab:lifetimes_eph_phph}
	\begin{tabularx}{\linewidth}{Y|C{1.3cm}|C{1.3cm}|C{1.3cm}|C{1.3cm}}
		& \multicolumn{4}{c}{\textbf{Lifetime / ps}} \\
		& \textbf{FT} & \textbf{FR} & \textbf{SA} & \textbf{IS} \\\hline
		sharp tip & 40.4 & 24.1 & 21.1 & 5.2 \\ 
		blunt tip & 30.0 & 4.1 & 15.2 & 9.0 \\
		slab adatom & 21.5 & 8.5 & 28.4 & 6.3 \\
		slab & 19.8 & 3.6 & 20.7 & 3.6 \\\hline
		CO on Cu(111)\cite{hirschmugl1994low} & - & 2.5\textsuperscript{*} & 2.5\textsuperscript{*} & 2.5\textsuperscript{*} \\
        CO on Cu(100)\cite{hirschmugl1994low} & - & 2.5\textsuperscript{*} & 1.9\textsuperscript{*} & 0.7\textsuperscript{*} \\
        CO on Cu(100)\cite{10.1063/1.471260} & 8 & - & - & - \\
	\end{tabularx}
\end{table}

\textbf{Kinetic energy decay.}
We can compare lifetimes determined through \textit{equilibrium correlation analysis} with those explicitly calculated from non-equilibrium \textit{kinetic energy decay} simulations (see Method section). Hereby, lifetimes are predicted by monitoring how the kinetic energy of an initially excited vibration mode decreases over time during a non-equilibrium MDEF simulation. We excite vibrations with a specific amount of energy and monitor the decay of the kinetic energy (Figure S18 in the Supporting Information). To determine the lifetimes, we assume that the vibrational modes behave like damped harmonic oscillators, where the envelope of the kinetic energy is assumed to exhibit an exponential decay. By including EPC using MDEF, total relaxation rates can be obtained. Overall, we observe that the kinetic energies of the CO vibrational modes indeed show an exponential decay (see Figure S18 in the Supporting Information).

\textbf{Additive relaxation rates.}
We sum two independent contributions to the vibrational relaxation rate: the PPC relaxation rates obtained from MD \textit{equilibrium correlation analysis}, and the EPC relaxation rates calculated in the quasi-static limit using first-order time-dependent perturbation theory. The EPC contribution calculated in this way is described in detail in \nameref{sec:electron_phonon_coupling}. The total relaxation rate $\gamma_\mathrm{tot}$ is the inverse of the lifetime $\gamma_\mathrm{x} = 1 / \tau_\mathrm{x}$ and can be approximated by summing the contributions from PPC $\gamma_\mathrm{ph-ph}$ (Table \ref{tab:lifetimes_phph}) and EPC $\gamma_\mathrm{e-ph}$ (Table \ref{tab:lifetimes_eph}):

\begin{equation}
    \gamma_\mathrm{tot} \approx \gamma_\mathrm{ph-ph} + \gamma_\mathrm{e-ph}
    \label{eq:relaxation_rate}
\end{equation}

Figure \ref{fig:lifetime_methods}b compares the three methods of determining total lifetimes (and relaxation rates). Additional technical details on the three methods are provided in Section S3 of the Supporting Information. Within the estimated uncertainty of our simulations, the results are in good agreement. All three methods yield lifetimes that exhibit similar trends across the different systems and modes. CO adsorption on a terrace leads to overall larger relaxation rates and smaller lifetimes compared to adsorption on adatoms or single atom tips. Relaxation rates exhibit a strong mode dependence, with the largest rate being an order of magnitude greater than the smallest. The numerical values in Figure \ref{fig:lifetime_methods}b can be found in Section S3 of the Supporting Information.

\textbf{Contributions to the relaxation rates.}
We decompose the total vibrational relaxation rate into EPC and EPP contributions to identify the dominant dissipation channel for each vibrational mode in each {\geom}. This is done using the \textit{kinetic energy decay} method. The phonon-phonon relaxation rates are obtained by monitoring the decay in kinetic energy during MD simulations (which do not include EPC). The electron-phonon relaxation rate is extracted from the MDEF trajectories. This is accomplished by integrating the electronic friction and random force components in equation \ref{eq:langevin_equation} along the trajectory.

Results for all {\geometries} are presented in Figure S22 in the Supporting Information and specifically for the slab {\geom} in Figure  \ref{fig:lifetime_methods}c. The relative contributions of PPC and EPC to the total relaxation rate are relatively independent of the {\geom} as shown in Figure S22 in the Supporting Information, but depend strongly on the vibrational mode. PPC primarily dominates the relaxation of FT and SA modes, whereas EPC predominantly influences the FR and IS modes.

Interestingly, the contributions from EPC and PPC do not fully add up to the total relaxation rate (see non-additive component in Figure \ref{fig:lifetime_methods}c) even when considering statistical uncertainties in the analysis. We observe a trend to underestimate the total rate, and the non-additivity is more pronounced in modes where EPC dominates (FR and IS modes). EPC and PPC may not be fully independent, as shown in Figures S24, S25, S26, and S27 in the Supporting Information. There, we find that PPC between the CO and the substrate vibrational modes increases if EPC is included in the equilibrium dynamics simulations. This result hints towards some level of non-additivity and a deviation from equation \ref{eq:relaxation_rate} when considering both effects simultaneously in the MDEF dynamics. Moreover, certain approximations taken in this work are expected to lead to more additivity. For example, we have assumed a configuration-independent electronic friction tensor. In reality, the friction tensor is configuration dependent and changes as the geometry is displaced along different vibration modes, likely contributing further to non-additivity.

\section*{Discussion}

\textbf{Vibrational modes and phonon-phonon coupling.} 
Remarkably, all phononic lifetimes fall within the same order of magnitude. In contrast, electron-phonon lifetimes vary by two orders of magnitude (IS vs. FT), highlighting a much stronger mode dependence compared to phonon-phonon interactions. Additionally, despite the significantly higher vibrational energy of the IS mode, its lifetime remains comparable to other modes, suggesting a surprising degree of energy independence in PPC. To understand this behaviour, we analyze the VDOS and coupling density. In the slab and slab-adatom systems, energetically more localized substrate vibrational states lead to reduced PPC compared to the sharp and blunt tips. This is counteracted by a larger number of vibrational states per volume (approximately three times as many) in the slab and slab-adatom systems. The FT mode has an energy comparable to the lowest-energy substrate vibrational modes, while the FR mode is close in energy to the highest-energy substrate modes. Coupling of the molecular vibrations is most efficient to the substrate vibrations with similar energy, as shown in Figure \ref{fig:coupling_dos_FT} and Supporting Information Figures S24, S25, S26, and S27 (Section S6.3). We can understand this in the following way: Energy conservation mandates that any newly excited vibrational mode arises from the creation or annihilation of $N$ other modes, following $E_\mathrm{new} = \sum_\mathrm{i}^\mathrm{N} n_\mathrm{i} E_\mathrm{i}$. As the number of participating modes increases, the efficiency of the process decreases. Because the FT and FR modes overlap energetically with the substrate modes, they can access efficient $N=1$ processes. However, individual substrate vibrations are strongly dependent on the surface shape. As a result, the phononic lifetimes of the FT and FR modes depend strongly on the surface shape, as shown in Table \ref{tab:lifetimes_phph}. The SA and IS modes are energetically relatively decoupled from the substrate modes compared to FT and FR. Therefore, their coupling of the substrate modes involves high-$N$ processes. Supporting Information Figures S26 and S27 show that the SA and IS modes couple equally to all surface modes. Consequently, the influence of surface geometry on specific surface vibrations is effectively averaged out, as shown in Table \ref{tab:lifetimes_phph}.

\textbf{Electron-phonon coupling.} 
We find small electron-phonon lifetimes for the FR and IS modes and large lifetimes for the FT and SA modes, in good agreement with experimental and theoretical literature.\cite{hirschmugl1994low, maurer2016ab, persson2004theory, lorente2005co} For each {\geom}, the electron-phonon lifetimes follow the trend of the vibrational frequencies of the respective modes, except for the FR mode. We can understand this behavior if we assume that the rate of EHP excitation is independent of the energies of the initial and final states. This corresponds to the wide-band, constant-coupling approximation, where one assumes a constant electronic DOS. In such a case, relaxation rates scale quadratically with the vibration energy. A similar explanation has been proposed by Persson.\cite{persson2004theory} As the vibrational energy increases, the available energy window for electronic transitions broadens, enabling more excitation pathways and thereby decreasing the lifetime. The vibrational energy of the IS mode is two orders of magnitude higher than that of the FT mode, consistent with the two-order-of-magnitude difference in their lifetimes. The FR mode deviates from this trend, with lifetimes comparable to the IS mode for the slab geometry. This can be explained by the large EPC reported by STM vibrational spectroscopy experiments for this mode.\cite{PhysRevB.60.R8525} Therefore, the EHP rate cannot be assumed to be independent of the involved energy states in this case.

Moving from the slab model to the sharp tip model, the lifetimes become larger. As Figure \ref{fig:dos} shows, the magnitude of the electronic PDOS becomes smaller with more corrugated {\geometries}. This leaves fewer electronic states available for electronic excitations. Consequently, EPC becomes less efficient, and the lifetime increases. However, the IS mode deviates from this trend. The available vibrational energy of the IS mode is approximately \SI{270}{\milli\electronvolt}, which is one order of magnitude larger than the energy of the SA and FR modes. The large vibrational energy of the IS mode increases the number of accessible electronic states for excitation. Even for the sharp tip, where the PDOS is highly localized, this energy range enables excitations between many more electronic states. Since the other modes carry significantly less energy, they are more strongly affected by the {\geom}: The lifetime of the FT mode increases by one order of magnitude between the slab and the sharp tip, making it two orders of magnitude longer than that of the IS mode.

\textbf{Combined energy dissipation mechanisms.}
We have determined total relaxation rates using three methods: \textit{equilibrium correlation analysis}, \textit{kinetic energy decay}, and \textit{additive relaxation rates}. Despite their significant conceptual difference, all three methods predict lifetimes that exhibit similar trends across the different systems and modes (see Figure \ref{fig:lifetime_methods}b). This demonstrates the robustness of our analysis. 

In Figure \ref{fig:lifetime_methods}c, the sum of phonon-phonon relaxation rates from MD and electron-phonon rates from MDEF simulations tends to be smaller than the total relaxation rates obtained directly from MDEF. For most modes, this difference lies within the uncertainties and is therefore strictly not statistically significant; it is in line with a comparison of the coupling densities with and without EPC: Coupling densities become larger if EPC is included in the simulations, indicating a weak non-additive effect where electron-phonon-mediated PPC increases phononic relaxation. However, the approximations in our approach likely underestimate this non-additivity. We assume a constant electronic friction tensor, whereas in reality, it is configuration-dependent. As the geometry is displaced along different vibration modes, the strength of EPC changes, likely enhancing the degree of non-additivity.

The observation from Figure \ref{fig:lifetime_methods}c that the relaxation rate of the FR (and IS) mode is dominated by EPC while the SA mode is not, is in good agreement with experiment. Asymmetric Fano line shapes have been reported for the FR mode, while symmetric Lorentzian line shapes have been measured for the SA mode.\cite{hirschmugl1994low} Fano line shapes typically indicate that a discrete vibrational mode interacts with a continuum of electronic states, indicating significant EPC. Conversely, a Lorentzian line shape suggests that PPC dominates the relaxation of the vibration mode.\cite{langreth1985energy}

However, it is important to note that the presence of EPC does not necessarily lead to a Fano line shape. Following Langreth,\cite{langreth1985energy} the line shape of a vibrational mode in the presence of EHP excitations is governed by the dynamic dipole $\mu$, with the asymmetry depending on the imaginary part of the equation:
\begin{equation}
    \mu = \left[ \mu_{\mathrm{ions}} + \Re({\mu_{\mathrm{el}}}) \right] \left[ 1 + i \frac{\Im({\mu_{\mathrm{el}}})}{\mu_{\mathrm{ions}} + \Re({\mu_{\mathrm{el}}})} \right]
    \label{eq:lineshape}
\end{equation}

Here $\mu_{\mathrm{ions}}$ and $\mu_{\mathrm{el}}$ are the ionic and electronic dipole, respectively. The electronic dipole consists of a real and an imaginary part. The imaginary part, $\Im({\mu_{\mathrm{el}}})$, emerges due to the creation of EHPs as adsorbate electrons and holes tunnel into the substrate. Since this tunneling process is not instantaneous, the oscillations of the electronic dipole get out of phase with the driving electric field.

The imaginary part of equation \ref{eq:lineshape} becomes large---giving rise to an asymmetric line shape---if the imaginary electronic dipole $\Im({\mu_{\mathrm{el}}})$ is large compared to the ionic dipole $\mu_{\mathrm{ions}}$. This occurs for modes with a dipole oriented parallel to the substrate---such as the FR mode---where screening by mirror charges significantly reduces the effective dipole strength. By contrast, modes with atomic motion perpendicular to the substrate, such as the SA and IS modes, exhibit a large ionic dipole. Subsequently, asymmetric line shapes would not be expected here, even if EPC is the dominant mechanism.

Our findings also inform the expected dynamic properties of CO on other coinage metals, as discussed in Section S4 of the Supporting Information. The greater mass mismatch between CO and Ag or Au, compared to Cu, reduces PPC. Moreover, the phonon density of states shifts to lower energies from Cu to Ag to Au (Figure S23 in the Supporting Information), leading to greater decoupling of the IS mode on Ag and Au than on Cu. Given the chemical similarity of the coinage metals, all of which are predominantly s-band conductors, this trend is expected to persist for different {\geometries}. Assuming comparable adsorption conditions, CO exhibits similar vibrational lifetimes on Cu, Ag, and Au.\cite{loncaric2019co, forsblom2007vibrational} However, due to weaker interactions with Ag and Au, EPC is significantly reduced.\cite{loncaric2019co} For adatoms or tip-like {\geometries}, trends observed on Cu likely extend to Ag and Au, although stronger interactions at low-coordination sites may facilitate shorter lifetimes relative to flat surfaces.

\textbf{Perspective and potential experimental validation.} 
Finally, vibrational lifetimes calculated with \textit{equilibrium correlation analysis} and \textit{kinetic energy decay} are in good agreement, which demonstrates that both are suitable and equivalent approaches. \textit{Equilibrium correlation analysis} can yield important information, such as coupling densities and phononic spectral functions. Our method for disentangling electron-phonon and phonon-phonon contributions to the vibrational dynamics of adsorbed molecules enables a better understanding of electronic and vibrational mass effects, as well as effects due to the local bonding environment. To connect theory with experiment, it is essential to identify techniques capable of validating mode-specific dissipation pathways. Most existing experiments for systems such as CO on Cu were conducted decades ago and are limited to flat surfaces.\cite{hirschmugl1994low, hirschmugl1995chemical} Data is lacking on low-energy modes, such as the FT. High resolution IR (infrared) and Raman spectroscopy allow for directly probing the dominant dissipation mechanisms via vibrational lineshapes: for vibrational modes with a small ionic dipole (such as those parallel to the surface where the dipole is screened), dominant EPC leads to characteristic Fano profiles (see Equation \ref{eq:lineshape}), while PPC results in Lorentzian lineshapes.\cite{langreth1985energy}

Tip-enhanced Raman spectroscopy (TERS) offers a powerful way to probe vibrational dynamics at the single-molecule level with atomic resolution, particularly when operated in the quantum tunneling regime of plasmonic enhancement.\cite{doi.org/10.1038/s41586-019-1059-9} By selectively tuning the excitation laser to match vibronic transitions, it becomes possible to achieve highly mode-specific resonance enhancement.\cite{doi.org/10.1038/s41467-024-51419-1} In STM-TERS, the application of a tunable bias voltage provides additional control: it can alter the tunneling gap and modulate the strength of EPC, which can be tracked via bias-dependent frequency shifts and intensity changes.\cite{doi:10.1021/acs.jpclett.8b01343} This could help to disentangle phonon-phonon and electron-phonon relaxation pathways and their contributions at specific adsorption sites.

To account for the adsorption site dependence (CO on different {\geometries}), experiments with spatial resolution are needed. STM-IETS (inelastic electron tunneling spectroscopy) can resolve low energy modes, is sensitive to IR and Raman inactive modes, and enables atomic spatial resolution.\cite{okabayashi2023dynamic} Crucially, it is highly sensitive to EPC: Electrons lose energy by exciting vibrations during tunneling, and the resulting peak intensity scales with the electron–phonon coupling strength.\cite{doi:10.1126/science.280.5370.1732, doi:10.1021/jp061590b} Meanwhile, the peaks are broadened by both EPC and PPC. By combining IETS with a different method of vibrational excitation, such as IR, which excites modes via changes in the dipole moment, it becomes possible to disentangle the dissipation mechanisms. For instance, a mode showing a large IR peak but a small IETS signal would indicate that PPC dominates, while a strong IETS response would indicate that EPC is significant.

Finally, our work can improve the interpretability of vibrational line widths and energy dissipation observed in AFM, IETS, TERS, or IR spectroscopy. The spectral function calculated through \textit{equilibrium correlation analysis} provides mode-specific linewidths and lifetimes, while the separation of electron–phonon and phonon–phonon contributions enables clearer interpretation of lineshapes and mode-specific enhancements, particularly in IR and TERS. By disentangling dissipation pathways, our method can facilitate cross-correlation between IR and IETS signals and help assign adsorption sites based on the {\geom}. Understanding which relaxation pathway dominates can enable experimental tuning. If the relaxation of a vibrational mode is primarily governed by EPC, adjusting electronic properties, such as the tip material, functionalization, or the bias voltage, can enhance signal detection and reveal details about the coupling. If PPC dominates, tuning temperature or using a different isotope (\textsuperscript{14}C or \textsuperscript{13}C instead of \textsuperscript{12}C in CO, for instance) can shift vibrational frequencies and affect the coupling. In scanning probe experiments and single-molecule manipulation, where vibrational excitation can induce molecular motion or reactions, understanding electron-phonon and phonon-phonon relaxation supports design conditions that favor specific molecular transformations. Similarly, in catalysis, targeting specific vibrational modes can selectively trigger reactions, enhancing the control and efficiency of chemical processes.

\section{Conclusion}

The vibrational dynamics of CO-decorated copper tips that range from atomically flat to maximally sharp exhibit pronounced vibrational mode specificity, driven by the interplay of EPC and PPC, as well as significant geometric effects. Lifetimes due to EPC differ by two orders of magnitude between modes due to vibrational energies, the symmetry of the mode, and the DOS. Overall, CO vibrations of flatter tip geometries exhibit stronger electron-phonon relaxation, due to the increased energetic delocalization of the local electronic DOS. However, the high-energy IS mode does not show this trend due to the large energy window to excite electrons. Phonon-phonon lifetimes also decrease as the surface becomes flatter, although the IS mode does not follow this trend. The FT, FR, and SA modes are coupled to the substrate vibrations, leading to significant geometry dependence, while the IS mode is decoupled from the substrate vibrations. The sum of individual electron–phonon and phonon–phonon relaxation rates tends to underestimate the total relaxation rate computed by the same method. This weak non-additivity arises from electron-phonon-mediated PPC, which enhances phonon-phonon interactions. Finally, the lifetimes obtained from \textit{equilibrium correlation analysis} and \textit{kinetic energy decay} are in good agreement and also agree well with experiment.

Future work will explore the role of non-diagonal electronic friction on the mode selectivity and surface dependence. Incorporating a fully configuration-dependent electronic friction tensor will enable a more accurate investigation of non-additivity.\cite{sachs_machine_2025} Finally, extending the current framework to account for zero-point energy contributions will further improve the fidelity of atomic-scale friction models.

\section{Methods}

\subsection{Modelling CO on different {\geometries}}

Electronic structure calculations are performed using DFT with the FHI-aims code,\cite{blum2009ab} employing periodic boundary conditions and a repeated-slab approach including a dipole correction. The sharp and blunt tip geometries are placed on a four-layer slab for better convergence. Convergence tests and further details are shown in Section S1 of the Supporting Information.

Commonly used methods for first-principles surface simulations, such as Generalized Gradient Approximation (GGA) functionals, fail to accurately describe CO adsorption on Cu and other metal surfaces.\cite{gajdovs2004co, gajdovs2005co, feibelman2001co} Due to the many-electron self-interaction error, the energies of the states responsible for bonding the CO molecule to the Cu surface are misrepresented. This results in an incorrect preference for the fcc-hollow site over the top site, which experimental studies have shown to be the correct adsorption site.\cite{vollmer2001determination} The Random Phase Approximation (RPA) has been proposed as a method to yield qualitatively correct adsorption sites, while the PBE0 functional shows a slight preference for the correct adsorption site.\cite{ren2009exploring} However, RPA calculations are unfeasible due to their high computational cost, while PBE0 is not ideal for modelling bulk metals since it tends to overestimate the delocalization of electrons, which can lead to an inaccurate description of metallic systems, particularly their electronic structure and density of states near the Fermi level.\cite{stroppa2008shortcomings, paier2007does} 

In our work, we use the HSE06 screened hybrid functional,\cite{heyd2003hybrid} which treats short-range interactions with exact exchange and reduces the amount of exact exchange at long range. This helps mitigate some of the issues, such as over-delocalization of electrons and poor convergence in calculations.\cite{janesko2009screened} We combine this function with the MBD-NL long-range dispersion correction\cite{hermann2020density}---a combination that was previously shown to provide accurate adsorption structure and energetics.\cite{stoodley2024structure} With this combination, we predict an accurate PES (see Table S1 in the Supporting Information). Geometry optimizations were performed for all systems with the four bottom layers frozen, prior to determining other properties (PDOS, EPC, PPC, etc.).

\subsection{Interatomic potential}

We use the MACE equivariant neural network potential\cite{batatia2022mace} and train it at the HSE06+MBD-NL level of theory, which accurately describes the potential energy surface of CO on Cu(111) as well as CO adsorbed on an adatom. Using a training set of 621 data points, the potential achieves an energy error of \SI{1.6}{\milli\electronvolt}/atom and a force error of \SI{15.9}{\milli\electronvolt\per\angstrom}, which was determined using k-fold cross-validation, where we split our dataset into five equally sized segments. The training data contains structures of all four {\geometries} (sharp tip, blunt tip, slab adatom, slab). Details regarding hyperparameters, training/test splits, and validation during dynamics simulations are shown in Section S2 in the Supporting Information.

\subsection{Vibration calculations}

Second-order force constants are calculated using finite displacements of \SI{0.0025}{\angstrom}. Energies and forces are determined using the MACE potential. During the vibration calculation, the bottom four layers of our systems are frozen. The second-order force constants are calculated for the top 10 mobile Cu layers as well as the CO molecule (approximately 500 atoms).

\subsection{Description of electron-phonon coupling}
\label{sec:description_of_electron_phonon_coupling}

In first-order perturbation theory, we can express the (3$\times$3) atomwise block of the electronic friction tensor $\mathbf{\Lambda}_\mathrm{i j}$ associated with atomic degrees of freedom $\mathbf{R}_\mathrm{i}$ and $\mathbf{R}_\mathrm{j}$ as follows:\cite{box2023ab, maurer2016ab}
\begin{equation}
\begin{aligned}
\mathbf{\Lambda}_\mathrm{i j}(\omega)= & 2 \pi \hbar \sum_\mathrm{\mathbf{k}, \nu, \nu^{\prime}>\nu} \left\langle \psi_\mathrm{\mathbf{k} \nu} \right|  \frac{\partial}{\partial \mathbf{R}_\mathrm{i}} \left| \psi_\mathrm{\mathbf{k} \nu^{\prime}} \right\rangle \left\langle \psi_\mathrm{\mathbf{k} \nu^{\prime}} \right| \frac{\partial}{\partial \mathbf{R}_\mathrm{j}} \left\rvert\, \psi_\mathrm{\mathbf{k} \nu} \right\rangle \\
& \times\left(f\left(\varepsilon_\mathrm{\mathbf{k} \nu}\right)-f\left(\varepsilon_\mathrm{\mathbf{k} \nu^{\prime}}\right)\right)\left(\varepsilon_\mathrm{\mathbf{k} \nu^{\prime}}-\varepsilon_\mathrm{\mathbf{k} \nu}\right) \\
& \times \delta\left(\varepsilon_\mathrm{\mathbf{k} \nu^{\prime}}-\varepsilon_\mathrm{\mathbf{k} \nu}-\hbar \omega\right)
\end{aligned}
\label{eq:electron_phonon_coupling}
\end{equation}

The electronic friction tensor associated with energy $\varepsilon = \hbar \omega$ of a given vibration mode is calculated by summing over all possible excitations between effective single particle Kohn–Sham states $\varepsilon_\mathrm{k \nu}$ and $\varepsilon_\mathrm{k \nu\prime}$. The indices $\nu$ and $\nu^{\prime}$ indicate the initial and final electronic state, respectively, and  $\mathbf{k}$ is the associated wave vector. The terms in the sum are given by the nonadiabatic coupling matrix elements (first row of Equation \ref{eq:electron_phonon_coupling}), the occupation factors $f(\varepsilon_\mathrm{\mathbf{k} \nu})$ and $f(\varepsilon_\mathrm{\mathbf{k} \nu^{\prime}})$ of the initial and final states (second row), and the conservation of energy (third row). Since we are dealing with electrons, the occupation factors follow the Fermi–Dirac distribution, which introduces the temperature dependence. The term $\hbar \omega$ accounts for the perturbing frequency of the vibration mode of interest. We calculate electronic friction with the electronic structure code FHI-aims.\cite{box2023ab} Convergence tests for electronic friction calculations can be found in Section 1.6 in the Supporting Information. We explicitly calculate electronic friction for C and O atoms. For the Cu atoms, we use isotropic electronic friction and a diagonal friction tensor. This friction value is calculated for a single atom in bulk copper (see Section S1.7 in the Supporting Information). EPC calculations are performed using the PBE\cite{perdew1996generalized} exchange-correlation functional, as has been done in previous simulations of CO on Cu substrates.\cite{maurer2016ab, box2023ab} Details can be found in Section S1.7 of the Supporting Information. We note that employing the HSE06 functional for electronic friction calculations in systems of this size is currently infeasible given available high-performance computing resources. Vibrational relaxation rates are determined from the friction tensor and the mass-weighted vibrational eigenmodes, $\tilde{\mathbf{u}}_{\alpha}$, at the relevant vibrational frequency $\omega_{\alpha}$:

\begin{align}\label{eq:broadening_with_friction}
    \gamma_{\alpha} = \hbar \left[ \tilde{\mathbf{u}}^T_{\alpha} \mathbf{\Lambda}(\hbar\omega_{\alpha})\tilde{\mathbf{u}}_{\alpha}  \right]
\end{align}

\subsection{Molecular dynamics}

We perform MD simulations using the atomic simulation environment (ASE).\cite{larsen2017atomic} A time step of \SI{0.1}{\femto\second} is employed. Since we are interested in the dynamics of a single CO molecule, we conduct MD simulations in an extended unit cell containing a single adsorbed CO molecule, shown in Figure \ref{fig:experimentalsetup}. The extended unit cell facilitates the back-folding of substrate phonon bands, enabling effective sampling of the phonon band structure.

To account for EPC, we use MDEF, where the dynamics are governed by the Langevin equation:\cite{head1995molecular}
\begin{equation}
    M_\mathrm{i} \Ddot{\mathbf{R}}_\mathrm{i} = - \frac{\partial V(\mathbf{R}) }{\partial \mathbf{R}_\mathrm{i}} - \sum_\mathrm{j} \mathbf{\Lambda}_\mathrm{ij} \dot{\mathbf{R}}_\mathrm{j} + \mathcal{R}_\mathrm{i}(T)
\label{eq:langevin_equation}
\end{equation}

Here, $V(\mathbf{R})$ represents the potential energy surface, which depends on the atomic positions $\mathbf{R}$. The term $\mathbf{\Lambda}_\mathrm{ij}$ denotes a (3$\times$3) atomwise block of the electronic friction tensor, while $\mathcal{R}_\mathrm{i}(T)$ is the random force vector responsible for maintaining detailed balance at a given electronic temperature $T$, as dictated by the second fluctuation-dissipation theorem.\cite{kubo1991statistical} For the dynamics, the electronic friction tensor is evaluated in the harmonic approximation at zero perturbation frequency, based on a fixed nuclear geometry. A comparison of lifetimes with and without a perturbing frequency can be found in Section S3.1 of the Supporting Information. During the calculations, the constant friction approximation is used for the electronic friction tensor, where the friction tensor values have been determined at the relaxed equilibrium geometry of the molecule on the surface. To determine relaxation rates and lifetimes from the MD and MDEF trajectories, we use three different methods:

\paragraph{\textbf{Equilibrium correlation analysis.}} For each system, we first perform equilibration at \SI{5}{\kelvin}, followed by computing a single, long MD or MDEF trajectory. To analyse the velocity time series, we use normal mode decomposition and determine the cross-spectrum using the Welch method. Thereby, the velocity time series is cut into multiple, partially overlapping segments. A window function---a Hann window in our case---is then applied to each segment before the cross-spectrum is calculated. The cross-spectra of the different segments are then averaged. This improves the smoothness of the cross-spectrum. The height and width of the peak(s) in the cross-spectrum are then determined by fitting a Lorentzian function. A detailed workflow is provided in Section S3.2 of the Supporting Information.
    
\paragraph{\textbf{Kinetic energy decay.}} For each system and vibrational mode, relaxation rates and lifetimes are extracted from ensemble averages over 64 MD or MDEF trajectories. We first equilibrate the system at \SI{5}{\kelvin}. Then, we excite the vibrational mode of interest by a displacement amplitude equivalent to \SI{50}{\milli\electronvolt} along the respective eigenvector. Using normal mode decomposition, we project the velocity time series onto the basis of the vibration modes. We subsequently monitor the temporal decay of the envelope of the ensemble-averaged kinetic energy. Assuming exponential decay---as would be the case for a damped harmonic oscillator---we determine relaxation rates via an exponential fit. See Section S3.3 of the Supporting Information for full details.

\paragraph{\textbf{Additive relaxation rates.}} Total relaxation rates are calculated by summing the individual relaxation rates associated with phonon-phonon and EPC. Phonon-phonon relaxation rates are derived from the inverse values listed in Table \ref{tab:lifetimes_phph}, which were determined through \textit{equilibrium correlation analysis}. Electron-phonon relaxation rates are obtained from the inverse values in Table \ref{tab:lifetimes_eph}, which were directly calculated from the electronic friction tensor. See Section S3.4 of the Supporting Information for the computational workflow.

\section{Acknowledgments}

The authors acknowledge financial support from the UKRI Future Leaders Fellowship program (MR/X023109/1), a UKRI Horizon grant (ERC StG, EP/X014088/1), and a UKRI Horizon grant (MSCA, EP/Y024923/1). We gratefully acknowledge valuable discussions with Nils Hertl and Connor Box. High-performance computing resources were provided via the Scientific Computing Research Technology Platform of the University of Warwick, the EPSRC-funded Materials Chemistry Consortium (EP/R029431/1, EP/X035859/1) and the UK Car-Parrinello consortium (EP/X035891/1) for the ARCHER2 UK National Supercomputing Service, the EPSRC-funded HPC Midlands+ computing centre for access to Sulis (EP/P020232/1), and EuroHPC Joint Undertaking and LUMI Consortium for access to LUMI (EHPC-EXT-2024E02-098).

\section{Data availability}

The data supporting this research has been uploaded to the NOMAD database at \href{10.17172/NOMAD/2025.04.29-1}{10.17172/NOMAD/2025.04.29-1}.

\section{Supporting information}

Extended description of electronic structure calculation settings and convergence tests; details of interatomic potential parametrization; technical information on lifetime calculations and tabulated numerical data; comparison of results with other coinage metals; additional analysis of molecular dynamics trajectories, including extended results on electron–phonon and phonon–phonon spectral functions.

\bibliography{bibliography}

\providecommand{\latin}[1]{#1}
\makeatletter
\providecommand{\doi}
  {\begingroup\let\do\@makeother\dospecials
  \catcode`\{=1 \catcode`\}=2 \doi@aux}
\providecommand{\doi@aux}[1]{\endgroup\texttt{#1}}
\makeatother
\providecommand*\mcitethebibliography{\thebibliography}
\csname @ifundefined\endcsname{endmcitethebibliography}
  {\let\endmcitethebibliography\endthebibliography}{}
\begin{mcitethebibliography}{68}
\providecommand*\natexlab[1]{#1}
\providecommand*\mciteSetBstSublistMode[1]{}
\providecommand*\mciteSetBstMaxWidthForm[2]{}
\providecommand*\mciteBstWouldAddEndPuncttrue
  {\def\EndOfBibitem{\unskip.}}
\providecommand*\mciteBstWouldAddEndPunctfalse
  {\let\EndOfBibitem\relax}
\providecommand*\mciteSetBstMidEndSepPunct[3]{}
\providecommand*\mciteSetBstSublistLabelBeginEnd[3]{}
\providecommand*\EndOfBibitem{}
\mciteSetBstSublistMode{f}
\mciteSetBstMaxWidthForm{subitem}{(\alph{mcitesubitemcount})}
\mciteSetBstSublistLabelBeginEnd
  {\mcitemaxwidthsubitemform\space}
  {\relax}
  {\relax}

\bibitem[Tully \latin{et~al.}(1993)Tully, Gomez, and
  Head-Gordon]{tully1993electronic}
Tully,~J.~C.; Gomez,~M.; Head-Gordon,~M. Electronic and phonon mechanisms of
  vibrational relaxation: CO on Cu (100). \emph{Journal of Vacuum Science \&
  Technology A: Vacuum, Surfaces, and Films} \textbf{1993}, \emph{11},
  1914--1920\relax
\mciteBstWouldAddEndPuncttrue
\mciteSetBstMidEndSepPunct{\mcitedefaultmidpunct}
{\mcitedefaultendpunct}{\mcitedefaultseppunct}\relax
\EndOfBibitem
\bibitem[Park and Salmeron(2014)Park, and Salmeron]{park2014fundamental}
Park,~J.~Y.; Salmeron,~M. Fundamental aspects of energy dissipation in
  friction. \emph{Chem. Rev.} \textbf{2014}, \emph{114}, 677--711\relax
\mciteBstWouldAddEndPuncttrue
\mciteSetBstMidEndSepPunct{\mcitedefaultmidpunct}
{\mcitedefaultendpunct}{\mcitedefaultseppunct}\relax
\EndOfBibitem
\bibitem[Okabayashi \latin{et~al.}(2023)Okabayashi, Frederiksen, Liebig, and
  Giessibl]{okabayashi2023dynamic}
Okabayashi,~N.; Frederiksen,~T.; Liebig,~A.; Giessibl,~F.~J. Dynamic friction
  unraveled by observing an unexpected intermediate state in controlled
  molecular manipulation. \emph{Phys. Rev. Lett.} \textbf{2023}, \emph{131},
  148001\relax
\mciteBstWouldAddEndPuncttrue
\mciteSetBstMidEndSepPunct{\mcitedefaultmidpunct}
{\mcitedefaultendpunct}{\mcitedefaultseppunct}\relax
\EndOfBibitem
\bibitem[Nam \latin{et~al.}(2024)Nam, Riegel, H{\"o}rmann, Hofmann, Gretz,
  Weymouth, and Giessibl]{nam2024exploring}
Nam,~S.; Riegel,~E.; H{\"o}rmann,~L.; Hofmann,~O.~T.; Gretz,~O.;
  Weymouth,~A.~J.; Giessibl,~F.~J. Exploring in-plane interactions beside an
  adsorbed molecule with lateral force microscopy. \emph{Proceedings of the
  National Academy of Sciences} \textbf{2024}, \emph{121}, e2311059120\relax
\mciteBstWouldAddEndPuncttrue
\mciteSetBstMidEndSepPunct{\mcitedefaultmidpunct}
{\mcitedefaultendpunct}{\mcitedefaultseppunct}\relax
\EndOfBibitem
\bibitem[Stipe \latin{et~al.}(1998)Stipe, Rezaei, and Ho]{stipe1998single}
Stipe,~B.~C.; Rezaei,~M.~A.; Ho,~W. Single-molecule vibrational spectroscopy
  and microscopy. \emph{Science} \textbf{1998}, \emph{280}, 1732--1735\relax
\mciteBstWouldAddEndPuncttrue
\mciteSetBstMidEndSepPunct{\mcitedefaultmidpunct}
{\mcitedefaultendpunct}{\mcitedefaultseppunct}\relax
\EndOfBibitem
\bibitem[Hirschmugl \latin{et~al.}(1994)Hirschmugl, Chabal, Hoffmann, and
  Williams]{hirschmugl1994low}
Hirschmugl,~C.; Chabal,~Y.; Hoffmann,~F.; Williams,~G. Low-frequency dynamics
  of CO/Cu breakdown of Born--Oppenheimer approximation. \emph{Journal of
  Vacuum Science \& Technology A: Vacuum, Surfaces, and Films} \textbf{1994},
  \emph{12}, 2229--2234\relax
\mciteBstWouldAddEndPuncttrue
\mciteSetBstMidEndSepPunct{\mcitedefaultmidpunct}
{\mcitedefaultendpunct}{\mcitedefaultseppunct}\relax
\EndOfBibitem
\bibitem[Hirschmugl and Williams(1995)Hirschmugl, and
  Williams]{hirschmugl1995chemical}
Hirschmugl,~C.~J.; Williams,~G.~P. Chemical shifts and coupling interactions
  for the bonding vibrational modes for CO/Cu (111) and (100) surfaces.
  \emph{Phys. Rev. B} \textbf{1995}, \emph{52}, 14177\relax
\mciteBstWouldAddEndPuncttrue
\mciteSetBstMidEndSepPunct{\mcitedefaultmidpunct}
{\mcitedefaultendpunct}{\mcitedefaultseppunct}\relax
\EndOfBibitem
\bibitem[Lauhon and Ho(1999)Lauhon, and Ho]{lauhon1999single}
Lauhon,~L.; Ho,~W. Single molecule thermal rotation and diffusion: Acetylene on
  Cu (001). \emph{J. Chem. Phys.} \textbf{1999}, \emph{111}, 5633--5636\relax
\mciteBstWouldAddEndPuncttrue
\mciteSetBstMidEndSepPunct{\mcitedefaultmidpunct}
{\mcitedefaultendpunct}{\mcitedefaultseppunct}\relax
\EndOfBibitem
\bibitem[Komeda \latin{et~al.}(2002)Komeda, Kim, Kawai, Persson, and
  Ueba]{komeda2002lateral}
Komeda,~T.; Kim,~Y.; Kawai,~M.; Persson,~B.; Ueba,~H. Lateral hopping of
  molecules induced by excitation of internal vibration mode. \emph{Science}
  \textbf{2002}, \emph{295}, 2055--2058\relax
\mciteBstWouldAddEndPuncttrue
\mciteSetBstMidEndSepPunct{\mcitedefaultmidpunct}
{\mcitedefaultendpunct}{\mcitedefaultseppunct}\relax
\EndOfBibitem
\bibitem[Nakamura \latin{et~al.}(2003)Nakamura, Mera, and
  Maeda]{nakamura2003hopping}
Nakamura,~Y.; Mera,~Y.; Maeda,~K. Hopping motion of chlorine atoms on Si (1 0
  0)-(2$\times$ 1) surfaces induced by carrier injection from scanning
  tunneling microscope tips. \emph{Surf. Sci.} \textbf{2003}, \emph{531},
  68--76\relax
\mciteBstWouldAddEndPuncttrue
\mciteSetBstMidEndSepPunct{\mcitedefaultmidpunct}
{\mcitedefaultendpunct}{\mcitedefaultseppunct}\relax
\EndOfBibitem
\bibitem[Braun \latin{et~al.}(1996)Braun, Graham, Hofmann, Silvestri, Toennies,
  and Witte]{braun1996he}
Braun,~J.; Graham,~A.~P.; Hofmann,~F.; Silvestri,~W.; Toennies,~J.~P.;
  Witte,~G. A He-atom scattering study of the frustrated translational mode of
  CO chemisorbed on defects on copper surfaces. \emph{J. Chem. Phys.}
  \textbf{1996}, \emph{105}, 3258--3263\relax
\mciteBstWouldAddEndPuncttrue
\mciteSetBstMidEndSepPunct{\mcitedefaultmidpunct}
{\mcitedefaultendpunct}{\mcitedefaultseppunct}\relax
\EndOfBibitem
\bibitem[Wruss \latin{et~al.}(2019)Wruss, H{\"o}rmann, and
  Hofmann]{wruss2019impact}
Wruss,~E.; H{\"o}rmann,~L.; Hofmann,~O.~T. Impact of surface defects on the
  charge transfer at inorganic/organic interfaces. \emph{The Journal of
  Physical Chemistry C} \textbf{2019}, \emph{123}, 7118--7124\relax
\mciteBstWouldAddEndPuncttrue
\mciteSetBstMidEndSepPunct{\mcitedefaultmidpunct}
{\mcitedefaultendpunct}{\mcitedefaultseppunct}\relax
\EndOfBibitem
\bibitem[Yudanov \latin{et~al.}(2003)Yudanov, Sahnoun, Neyman, R{\"o}sch,
  Hoffmann, Schauermann, Johanek, Unterhalt, Rupprechter, Libuda,
  \latin{et~al.} others]{yudanov2003co}
Yudanov,~I.~V.; Sahnoun,~R.; Neyman,~K.~M.; R{\"o}sch,~N.; Hoffmann,~J.;
  Schauermann,~S.; Johanek,~V.; Unterhalt,~H.; Rupprechter,~G.; Libuda,~J.;
  others CO adsorption on Pd nanoparticles: Density functional and vibrational
  spectroscopy studies. \emph{The Journal of Physical Chemistry B}
  \textbf{2003}, \emph{107}, 255--264\relax
\mciteBstWouldAddEndPuncttrue
\mciteSetBstMidEndSepPunct{\mcitedefaultmidpunct}
{\mcitedefaultendpunct}{\mcitedefaultseppunct}\relax
\EndOfBibitem
\bibitem[Sader \latin{et~al.}(2012)Sader, Sanelli, Adamson, Monty, Wei,
  Crawford, Friend, Marusic, Mulvaney, and Bieske]{sader2012spring}
Sader,~J.~E.; Sanelli,~J.~A.; Adamson,~B.~D.; Monty,~J.~P.; Wei,~X.;
  Crawford,~S.~A.; Friend,~J.~R.; Marusic,~I.; Mulvaney,~P.; Bieske,~E.~J.
  Spring constant calibration of atomic force microscope cantilevers of
  arbitrary shape. \emph{Rev. Sci. Instrum.} \textbf{2012}, \emph{83}\relax
\mciteBstWouldAddEndPuncttrue
\mciteSetBstMidEndSepPunct{\mcitedefaultmidpunct}
{\mcitedefaultendpunct}{\mcitedefaultseppunct}\relax
\EndOfBibitem
\bibitem[Walton(1977)]{walton1977triboluminescence}
Walton,~A.~J. Triboluminescence. \emph{Adv. Phys.} \textbf{1977}, \emph{26},
  887--948\relax
\mciteBstWouldAddEndPuncttrue
\mciteSetBstMidEndSepPunct{\mcitedefaultmidpunct}
{\mcitedefaultendpunct}{\mcitedefaultseppunct}\relax
\EndOfBibitem
\bibitem[Dai and Ho(1995)Dai, and Ho]{dai1995laser}
Dai,~H.-L.; Ho,~W. \emph{Laser Spectroscopy And Photochemistry On Metal
  Surfaces (In 2 Parts)-Part 2}; World Scientific, 1995; Vol.~5\relax
\mciteBstWouldAddEndPuncttrue
\mciteSetBstMidEndSepPunct{\mcitedefaultmidpunct}
{\mcitedefaultendpunct}{\mcitedefaultseppunct}\relax
\EndOfBibitem
\bibitem[Ge \latin{et~al.}(2018)Ge, Rudshteyn, Zhu, Maurer, Batista, and
  Lian]{ge2018electron}
Ge,~A.; Rudshteyn,~B.; Zhu,~J.; Maurer,~R.~J.; Batista,~V.~S.; Lian,~T.
  Electron--hole-pair-induced vibrational energy relaxation of rhenium
  catalysts on gold surfaces. \emph{J. Phys. Chem. Lett.} \textbf{2018},
  \emph{9}, 406--412\relax
\mciteBstWouldAddEndPuncttrue
\mciteSetBstMidEndSepPunct{\mcitedefaultmidpunct}
{\mcitedefaultendpunct}{\mcitedefaultseppunct}\relax
\EndOfBibitem
\bibitem[Hollins and Pritchard(1979)Hollins, and
  Pritchard]{hollins1979interactions}
Hollins,~P.; Pritchard,~J. Interactions of CO molecules adsorbed on Cu (111).
  \emph{Surf. Sci.} \textbf{1979}, \emph{89}, 486--495\relax
\mciteBstWouldAddEndPuncttrue
\mciteSetBstMidEndSepPunct{\mcitedefaultmidpunct}
{\mcitedefaultendpunct}{\mcitedefaultseppunct}\relax
\EndOfBibitem
\bibitem[Weymouth \latin{et~al.}(2014)Weymouth, Hofmann, and
  Giessibl]{weymouth2014quantifying}
Weymouth,~A.~J.; Hofmann,~T.; Giessibl,~F.~J. Quantifying molecular stiffness
  and interaction with lateral force microscopy. \emph{Science} \textbf{2014},
  \emph{343}, 1120--1122\relax
\mciteBstWouldAddEndPuncttrue
\mciteSetBstMidEndSepPunct{\mcitedefaultmidpunct}
{\mcitedefaultendpunct}{\mcitedefaultseppunct}\relax
\EndOfBibitem
\bibitem[Gameel \latin{et~al.}(2018)Gameel, Sharafeldin, Abourayya, Biby, and
  Allam]{gameel2018unveiling}
Gameel,~K.~M.; Sharafeldin,~I.~M.; Abourayya,~A.~U.; Biby,~A.~H.; Allam,~N.~K.
  Unveiling CO adsorption on Cu surfaces: new insights from molecular orbital
  principles. \emph{Physical Chemistry Chemical Physics} \textbf{2018},
  \emph{20}, 25892--25900\relax
\mciteBstWouldAddEndPuncttrue
\mciteSetBstMidEndSepPunct{\mcitedefaultmidpunct}
{\mcitedefaultendpunct}{\mcitedefaultseppunct}\relax
\EndOfBibitem
\bibitem[Gross \latin{et~al.}(2009)Gross, Mohn, Moll, Liljeroth, and
  Meyer]{gross2009chemical}
Gross,~L.; Mohn,~F.; Moll,~N.; Liljeroth,~P.; Meyer,~G. The chemical structure
  of a molecule resolved by atomic force microscopy. \emph{Science}
  \textbf{2009}, \emph{325}, 1110--1114\relax
\mciteBstWouldAddEndPuncttrue
\mciteSetBstMidEndSepPunct{\mcitedefaultmidpunct}
{\mcitedefaultendpunct}{\mcitedefaultseppunct}\relax
\EndOfBibitem
\bibitem[Hammer and Nørskov(1995)Hammer, and Nørskov]{HAMMER1995211}
Hammer,~B.; Nørskov,~J. Electronic factors determining the reactivity of metal
  surfaces. \emph{Surface Science} \textbf{1995}, \emph{343}, 211--220\relax
\mciteBstWouldAddEndPuncttrue
\mciteSetBstMidEndSepPunct{\mcitedefaultmidpunct}
{\mcitedefaultendpunct}{\mcitedefaultseppunct}\relax
\EndOfBibitem
\bibitem[Maurer \latin{et~al.}(2016)Maurer, Askerka, Batista, and
  Tully]{maurer2016ab}
Maurer,~R.~J.; Askerka,~M.; Batista,~V.~S.; Tully,~J.~C. Ab initio tensorial
  electronic friction for molecules on metal surfaces: Nonadiabatic vibrational
  relaxation. \emph{Phys. Rev. B} \textbf{2016}, \emph{94}, 115432\relax
\mciteBstWouldAddEndPuncttrue
\mciteSetBstMidEndSepPunct{\mcitedefaultmidpunct}
{\mcitedefaultendpunct}{\mcitedefaultseppunct}\relax
\EndOfBibitem
\bibitem[Lorente and Ueba(2005)Lorente, and Ueba]{lorente2005co}
Lorente,~N.; Ueba,~H. CO dynamics induced by tunneling electrons: differences
  on Cu (110) and Ag (110). \emph{Eur. Phys. J. D} \textbf{2005}, \emph{35},
  341--348\relax
\mciteBstWouldAddEndPuncttrue
\mciteSetBstMidEndSepPunct{\mcitedefaultmidpunct}
{\mcitedefaultendpunct}{\mcitedefaultseppunct}\relax
\EndOfBibitem
\bibitem[Persson(2004)]{persson2004theory}
Persson,~M. Theory of elastic and inelastic tunnelling microscopy and
  spectroscopy: CO on Cu revisited. \emph{Philos. Trans. R. Soc. Lond. A}
  \textbf{2004}, \emph{362}, 1173--1183\relax
\mciteBstWouldAddEndPuncttrue
\mciteSetBstMidEndSepPunct{\mcitedefaultmidpunct}
{\mcitedefaultendpunct}{\mcitedefaultseppunct}\relax
\EndOfBibitem
\bibitem[Gajdo{\v{s}} \latin{et~al.}(2004)Gajdo{\v{s}}, Eichler, and
  Hafner]{gajdovs2004co}
Gajdo{\v{s}},~M.; Eichler,~A.; Hafner,~J. CO adsorption on close-packed
  transition and noble metal surfaces: trends from ab initio calculations.
  \emph{J. Condens. Matter Phys.} \textbf{2004}, \emph{16}, 1141\relax
\mciteBstWouldAddEndPuncttrue
\mciteSetBstMidEndSepPunct{\mcitedefaultmidpunct}
{\mcitedefaultendpunct}{\mcitedefaultseppunct}\relax
\EndOfBibitem
\bibitem[Gajdo{\v{s}} and Hafner(2005)Gajdo{\v{s}}, and Hafner]{gajdovs2005co}
Gajdo{\v{s}},~M.; Hafner,~J. CO adsorption on Cu (1 1 1) and Cu (0 0 1)
  surfaces: Improving site preference in DFT calculations. \emph{Surf. Sci.}
  \textbf{2005}, \emph{590}, 117--126\relax
\mciteBstWouldAddEndPuncttrue
\mciteSetBstMidEndSepPunct{\mcitedefaultmidpunct}
{\mcitedefaultendpunct}{\mcitedefaultseppunct}\relax
\EndOfBibitem
\bibitem[Feibelman \latin{et~al.}(2001)Feibelman, Hammer, N{\o}rskov, Wagner,
  Scheffler, Stumpf, Watwe, and Dumesic]{feibelman2001co}
Feibelman,~P.~J.; Hammer,~B.; N{\o}rskov,~J.~K.; Wagner,~F.; Scheffler,~M.;
  Stumpf,~R.; Watwe,~R.; Dumesic,~J. The co/pt (111) puzzle. \emph{The Journal
  of Physical Chemistry B} \textbf{2001}, \emph{105}, 4018--4025\relax
\mciteBstWouldAddEndPuncttrue
\mciteSetBstMidEndSepPunct{\mcitedefaultmidpunct}
{\mcitedefaultendpunct}{\mcitedefaultseppunct}\relax
\EndOfBibitem
\bibitem[Heyd \latin{et~al.}(2003)Heyd, Scuseria, and
  Ernzerhof]{heyd2003hybrid}
Heyd,~J.; Scuseria,~G.~E.; Ernzerhof,~M. Hybrid functionals based on a screened
  Coulomb potential. \emph{J. Chem. Phys.} \textbf{2003}, \emph{118},
  8207--8215\relax
\mciteBstWouldAddEndPuncttrue
\mciteSetBstMidEndSepPunct{\mcitedefaultmidpunct}
{\mcitedefaultendpunct}{\mcitedefaultseppunct}\relax
\EndOfBibitem
\bibitem[Hermann and Tkatchenko(2020)Hermann, and
  Tkatchenko]{hermann2020density}
Hermann,~J.; Tkatchenko,~A. Density functional model for van der Waals
  interactions: Unifying many-body atomic approaches with nonlocal functionals.
  \emph{Phys. Rev. Lett.} \textbf{2020}, \emph{124}, 146401\relax
\mciteBstWouldAddEndPuncttrue
\mciteSetBstMidEndSepPunct{\mcitedefaultmidpunct}
{\mcitedefaultendpunct}{\mcitedefaultseppunct}\relax
\EndOfBibitem
\bibitem[Tully(1990)]{tully1990molecular}
Tully,~J.~C. Molecular dynamics with electronic transitions. \emph{J. Chem.
  Phys.} \textbf{1990}, \emph{93}, 1061--1071\relax
\mciteBstWouldAddEndPuncttrue
\mciteSetBstMidEndSepPunct{\mcitedefaultmidpunct}
{\mcitedefaultendpunct}{\mcitedefaultseppunct}\relax
\EndOfBibitem
\bibitem[Hapala \latin{et~al.}(2014)Hapala, Temirov, Tautz, and
  Jel{\'\i}nek]{hapala2014origin}
Hapala,~P.; Temirov,~R.; Tautz,~F.~S.; Jel{\'\i}nek,~P. Origin of
  high-resolution IETS-STM images of organic molecules with functionalized
  tips. \emph{Phys. Rev. Lett.} \textbf{2014}, \emph{113}, 226101\relax
\mciteBstWouldAddEndPuncttrue
\mciteSetBstMidEndSepPunct{\mcitedefaultmidpunct}
{\mcitedefaultendpunct}{\mcitedefaultseppunct}\relax
\EndOfBibitem
\bibitem[Peller \latin{et~al.}(2021)Peller, Roelcke, Kastner, Buchner, Neef,
  Hayes, Bonaf{\'e}, Sidler, Ruggenthaler, Rubio, \latin{et~al.}
  others]{peller2021quantitative}
Peller,~D.; Roelcke,~C.; Kastner,~L.~Z.; Buchner,~T.; Neef,~A.; Hayes,~J.;
  Bonaf{\'e},~F.; Sidler,~D.; Ruggenthaler,~M.; Rubio,~A.; others Quantitative
  sampling of atomic-scale electromagnetic waveforms. \emph{Nat. Photonics}
  \textbf{2021}, \emph{15}, 143--147\relax
\mciteBstWouldAddEndPuncttrue
\mciteSetBstMidEndSepPunct{\mcitedefaultmidpunct}
{\mcitedefaultendpunct}{\mcitedefaultseppunct}\relax
\EndOfBibitem
\bibitem[Batatia \latin{et~al.}(2022)Batatia, Kovacs, Simm, Ortner, and
  Cs{\'a}nyi]{batatia2022mace}
Batatia,~I.; Kovacs,~D.~P.; Simm,~G.; Ortner,~C.; Cs{\'a}nyi,~G. MACE: Higher
  order equivariant message passing neural networks for fast and accurate force
  fields. \emph{Advances in Neural Information Processing Systems}
  \textbf{2022}, \emph{35}, 11423--11436\relax
\mciteBstWouldAddEndPuncttrue
\mciteSetBstMidEndSepPunct{\mcitedefaultmidpunct}
{\mcitedefaultendpunct}{\mcitedefaultseppunct}\relax
\EndOfBibitem
\bibitem[Sun \latin{et~al.}(2014)Sun, Zhang, and Wentzcovitch]{sun2014dynamic}
Sun,~T.; Zhang,~D.-B.; Wentzcovitch,~R.~M. Dynamic stabilization of cubic Ca Si
  O 3 perovskite at high temperatures and pressures from ab initio molecular
  dynamics. \emph{Phys. Rev. B} \textbf{2014}, \emph{89}, 094109\relax
\mciteBstWouldAddEndPuncttrue
\mciteSetBstMidEndSepPunct{\mcitedefaultmidpunct}
{\mcitedefaultendpunct}{\mcitedefaultseppunct}\relax
\EndOfBibitem
\bibitem[Liu \latin{et~al.}(2024)Liu, Yang, Zhu, Li, Li, Zhai, Song, Yang, and
  Li]{doi:10.1021/jacsau.4c00775}
Liu,~J.; Yang,~J.; Zhu,~G.; Li,~J.; Li,~Y.; Zhai,~Y.; Song,~H.; Yang,~Y.;
  Li,~H. Revealing the Ultrafast Energy Transfer Pathways in Energetic
  Materials: Time-Dependent and Quantum State-Resolved. \emph{JACS Au}
  \textbf{2024}, \emph{4}, 4455--4465\relax
\mciteBstWouldAddEndPuncttrue
\mciteSetBstMidEndSepPunct{\mcitedefaultmidpunct}
{\mcitedefaultendpunct}{\mcitedefaultseppunct}\relax
\EndOfBibitem
\bibitem[Arnolds and Bonn(2010)Arnolds, and Bonn]{ARNOLDS201045}
Arnolds,~H.; Bonn,~M. Ultrafast surface vibrational dynamics. \emph{Surf. Sci.
  Rep.} \textbf{2010}, \emph{65}, 45--66\relax
\mciteBstWouldAddEndPuncttrue
\mciteSetBstMidEndSepPunct{\mcitedefaultmidpunct}
{\mcitedefaultendpunct}{\mcitedefaultseppunct}\relax
\EndOfBibitem
\bibitem[Box \latin{et~al.}(2023)Box, Stark, and Maurer]{box2023ab}
Box,~C.~L.; Stark,~W.~G.; Maurer,~R.~J. Ab initio calculation of
  electron-phonon linewidths and molecular dynamics with electronic friction at
  metal surfaces with numeric atom-centred orbitals. \emph{Electron. Struct.}
  \textbf{2023}, \emph{5}, 035005\relax
\mciteBstWouldAddEndPuncttrue
\mciteSetBstMidEndSepPunct{\mcitedefaultmidpunct}
{\mcitedefaultendpunct}{\mcitedefaultseppunct}\relax
\EndOfBibitem
\bibitem[Maurer \latin{et~al.}(2017)Maurer, Jiang, Guo, and
  Tully]{maurer2017mode}
Maurer,~R.~J.; Jiang,~B.; Guo,~H.; Tully,~J.~C. Mode specific electronic
  friction in dissociative chemisorption on metal surfaces: H 2 on Ag (111).
  \emph{Phys. Rev. Lett.} \textbf{2017}, \emph{118}, 256001\relax
\mciteBstWouldAddEndPuncttrue
\mciteSetBstMidEndSepPunct{\mcitedefaultmidpunct}
{\mcitedefaultendpunct}{\mcitedefaultseppunct}\relax
\EndOfBibitem
\bibitem[Box \latin{et~al.}(2020)Box, Zhang, Yin, Jiang, and
  Maurer]{box2020determining}
Box,~C.~L.; Zhang,~Y.; Yin,~R.; Jiang,~B.; Maurer,~R.~J. Determining the effect
  of hot electron dissipation on molecular scattering experiments at metal
  surfaces. \emph{JACS Au} \textbf{2020}, \emph{1}, 164--173\relax
\mciteBstWouldAddEndPuncttrue
\mciteSetBstMidEndSepPunct{\mcitedefaultmidpunct}
{\mcitedefaultendpunct}{\mcitedefaultseppunct}\relax
\EndOfBibitem
\bibitem[Novko \latin{et~al.}(2018)Novko, Alducin, and
  Juaristi]{PhysRevLett.120.156804}
Novko,~D.; Alducin,~M.; Juaristi,~J.~I. Electron-Mediated Phonon-Phonon
  Coupling Drives the Vibrational Relaxation of CO on Cu(100). \emph{Phys. Rev.
  Lett.} \textbf{2018}, \emph{120}, 156804\relax
\mciteBstWouldAddEndPuncttrue
\mciteSetBstMidEndSepPunct{\mcitedefaultmidpunct}
{\mcitedefaultendpunct}{\mcitedefaultseppunct}\relax
\EndOfBibitem
\bibitem[Askerka \latin{et~al.}(2016)Askerka, Maurer, Batista, and
  Tully]{askerka2016role}
Askerka,~M.; Maurer,~R.~J.; Batista,~V.~S.; Tully,~J.~C. Role of tensorial
  electronic friction in energy transfer at metal surfaces. \emph{Phys. Rev.
  Lett.} \textbf{2016}, \emph{116}, 217601\relax
\mciteBstWouldAddEndPuncttrue
\mciteSetBstMidEndSepPunct{\mcitedefaultmidpunct}
{\mcitedefaultendpunct}{\mcitedefaultseppunct}\relax
\EndOfBibitem
\bibitem[Janke \latin{et~al.}(2015)Janke, Auerbach, Wodtke, and
  Kandratsenka]{10.1063/1.4931669}
Janke,~S.~M.; Auerbach,~D.~J.; Wodtke,~A.~M.; Kandratsenka,~A. An accurate
  full-dimensional potential energy surface for H–Au(111): Importance of
  nonadiabatic electronic excitation in energy transfer and adsorption.
  \emph{J. Chem. Phys.} \textbf{2015}, \emph{143}, 124708\relax
\mciteBstWouldAddEndPuncttrue
\mciteSetBstMidEndSepPunct{\mcitedefaultmidpunct}
{\mcitedefaultendpunct}{\mcitedefaultseppunct}\relax
\EndOfBibitem
\bibitem[Spiering \latin{et~al.}(2019)Spiering, Shakouri, Behler, Kroes, and
  Meyer]{spiering2019orbital}
Spiering,~P.; Shakouri,~K.; Behler,~J.; Kroes,~G.-J.; Meyer,~J.
  Orbital-dependent electronic friction significantly affects the description
  of reactive scattering of N2 from Ru (0001). \emph{J. Phys. Chem. Lett.}
  \textbf{2019}, \emph{10}, 2957--2962\relax
\mciteBstWouldAddEndPuncttrue
\mciteSetBstMidEndSepPunct{\mcitedefaultmidpunct}
{\mcitedefaultendpunct}{\mcitedefaultseppunct}\relax
\EndOfBibitem
\bibitem[Graham \latin{et~al.}(1996)Graham, Hofmann, and
  Toennies]{graham1996observation}
Graham,~A.; Hofmann,~F.; Toennies,~J.~P. Observation of the broadening and
  shift of the frustrated translation vibrational mode of CO on Cu (001) by
  high resolution helium atom scattering. \emph{J. Chem. Phys.} \textbf{1996},
  \emph{104}, 5311--5316\relax
\mciteBstWouldAddEndPuncttrue
\mciteSetBstMidEndSepPunct{\mcitedefaultmidpunct}
{\mcitedefaultendpunct}{\mcitedefaultseppunct}\relax
\EndOfBibitem
\bibitem[Graham \latin{et~al.}(1996)Graham, Hofmann, and
  Toennies]{10.1063/1.471260}
Graham,~A.; Hofmann,~F.; Toennies,~J.~P. Observation of the broadening and
  shift of the frustrated translation vibrational mode of CO on Cu(001) by high
  resolution helium atom scattering. \emph{J. Chem. Phys.} \textbf{1996},
  \emph{104}, 5311--5316\relax
\mciteBstWouldAddEndPuncttrue
\mciteSetBstMidEndSepPunct{\mcitedefaultmidpunct}
{\mcitedefaultendpunct}{\mcitedefaultseppunct}\relax
\EndOfBibitem
\bibitem[Lauhon and Ho(1999)Lauhon, and Ho]{PhysRevB.60.R8525}
Lauhon,~L.~J.; Ho,~W. Single-molecule vibrational spectroscopy and microscopy:
  CO on Cu(001) and Cu(110). \emph{Phys. Rev. B} \textbf{1999}, \emph{60},
  R8525--R8528\relax
\mciteBstWouldAddEndPuncttrue
\mciteSetBstMidEndSepPunct{\mcitedefaultmidpunct}
{\mcitedefaultendpunct}{\mcitedefaultseppunct}\relax
\EndOfBibitem
\bibitem[Langreth(1985)]{langreth1985energy}
Langreth,~D.~C. Energy transfer at surfaces: asymmetric line shapes and the
  electron-hole-pair mechanism. \emph{Phys. Rev. Lett.} \textbf{1985},
  \emph{54}, 126\relax
\mciteBstWouldAddEndPuncttrue
\mciteSetBstMidEndSepPunct{\mcitedefaultmidpunct}
{\mcitedefaultendpunct}{\mcitedefaultseppunct}\relax
\EndOfBibitem
\bibitem[Lon{\v{c}}ari{\'c} \latin{et~al.}(2019)Lon{\v{c}}ari{\'c}, Alducin,
  Juaristi, and Novko]{loncaric2019co}
Lon{\v{c}}ari{\'c},~I.; Alducin,~M.; Juaristi,~J.~I.; Novko,~D. CO Stretch
  Vibration Lives Long on Au(111). \emph{The Journal of Physical Chemistry
  Letters} \textbf{2019}, \emph{10}, 1043--1047\relax
\mciteBstWouldAddEndPuncttrue
\mciteSetBstMidEndSepPunct{\mcitedefaultmidpunct}
{\mcitedefaultendpunct}{\mcitedefaultseppunct}\relax
\EndOfBibitem
\bibitem[Forsblom and Persson(2007)Forsblom, and
  Persson]{forsblom2007vibrational}
Forsblom,~M.; Persson,~M. Vibrational lifetimes of cyanide and carbon monoxide
  on noble and transition metal surfaces. \emph{The Journal of chemical
  physics} \textbf{2007}, \emph{127}\relax
\mciteBstWouldAddEndPuncttrue
\mciteSetBstMidEndSepPunct{\mcitedefaultmidpunct}
{\mcitedefaultendpunct}{\mcitedefaultseppunct}\relax
\EndOfBibitem
\bibitem[Lee \latin{et~al.}(2019)Lee, Crampton, Tallarida, and
  Apkarian]{doi.org/10.1038/s41586-019-1059-9}
Lee,~J.; Crampton,~K.~T.; Tallarida,~N.; Apkarian,~V.~A. Visualizing
  vibrational normal modes of a single molecule with atomically confined light.
  \emph{Nature} \textbf{2019}, \emph{568}, 78--82\relax
\mciteBstWouldAddEndPuncttrue
\mciteSetBstMidEndSepPunct{\mcitedefaultmidpunct}
{\mcitedefaultendpunct}{\mcitedefaultseppunct}\relax
\EndOfBibitem
\bibitem[Luo \latin{et~al.}(2024)Luo, Sheng, Pisarra, Martin-Jimenez, Martin,
  Kern, and Garg]{doi.org/10.1038/s41467-024-51419-1}
Luo,~Y.; Sheng,~S.; Pisarra,~M.; Martin-Jimenez,~A.; Martin,~F.; Kern,~K.;
  Garg,~M. Selective excitation of vibrations in a single molecule.
  \emph{Nature Communications} \textbf{2024}, \emph{15}, 6983\relax
\mciteBstWouldAddEndPuncttrue
\mciteSetBstMidEndSepPunct{\mcitedefaultmidpunct}
{\mcitedefaultendpunct}{\mcitedefaultseppunct}\relax
\EndOfBibitem
\bibitem[Gieseking \latin{et~al.}(2018)Gieseking, Lee, Tallarida, Apkarian, and
  Schatz]{doi:10.1021/acs.jpclett.8b01343}
Gieseking,~R. L.~M.; Lee,~J.; Tallarida,~N.; Apkarian,~V.~A.; Schatz,~G.~C.
  Bias-Dependent Chemical Enhancement and Nonclassical Stark Effect in
  Tip-Enhanced Raman Spectromicroscopy of CO-Terminated Ag Tips. \emph{The
  Journal of Physical Chemistry Letters} \textbf{2018}, \emph{9}, 3074--3080,
  PMID: 29782171\relax
\mciteBstWouldAddEndPuncttrue
\mciteSetBstMidEndSepPunct{\mcitedefaultmidpunct}
{\mcitedefaultendpunct}{\mcitedefaultseppunct}\relax
\EndOfBibitem
\bibitem[Stipe \latin{et~al.}(1998)Stipe, Rezaei, and
  Ho]{doi:10.1126/science.280.5370.1732}
Stipe,~B.~C.; Rezaei,~M.~A.; Ho,~W. Single-Molecule Vibrational Spectroscopy
  and Microscopy. \emph{Science} \textbf{1998}, \emph{280}, 1732--1735\relax
\mciteBstWouldAddEndPuncttrue
\mciteSetBstMidEndSepPunct{\mcitedefaultmidpunct}
{\mcitedefaultendpunct}{\mcitedefaultseppunct}\relax
\EndOfBibitem
\bibitem[Maddox \latin{et~al.}(2006)Maddox, Harbola, Liu, Silien, Ho, Bazan,
  and Mukamel]{doi:10.1021/jp061590b}
Maddox,~J.~B.; Harbola,~U.; Liu,~N.; Silien,~C.; Ho,~W.; Bazan,~G.~C.;
  Mukamel,~S. Simulation of Single Molecule Inelastic Electron Tunneling
  Signals in Paraphenylene-Vinylene Oligomers and
  Distyrylbenzene[2.2]paracyclophanes. \emph{The Journal of Physical Chemistry
  A} \textbf{2006}, \emph{110}, 6329--6338, PMID: 16686469\relax
\mciteBstWouldAddEndPuncttrue
\mciteSetBstMidEndSepPunct{\mcitedefaultmidpunct}
{\mcitedefaultendpunct}{\mcitedefaultseppunct}\relax
\EndOfBibitem
\bibitem[Sachs \latin{et~al.}(2025)Sachs, Stark, Maurer, and
  Ortner]{sachs_machine_2025}
Sachs,~M.; Stark,~W.~G.; Maurer,~R.~J.; Ortner,~C. Machine learning
  configuration-dependent friction tensors in {Langevin} heatbaths. \emph{Mach.
  learn.: Sci. Technol.} \textbf{2025}, \emph{6}, 015016, Publisher: IOP
  Publishing\relax
\mciteBstWouldAddEndPuncttrue
\mciteSetBstMidEndSepPunct{\mcitedefaultmidpunct}
{\mcitedefaultendpunct}{\mcitedefaultseppunct}\relax
\EndOfBibitem
\bibitem[Blum \latin{et~al.}(2009)Blum, Gehrke, Hanke, Havu, Havu, Ren, Reuter,
  and Scheffler]{blum2009ab}
Blum,~V.; Gehrke,~R.; Hanke,~F.; Havu,~P.; Havu,~V.; Ren,~X.; Reuter,~K.;
  Scheffler,~M. Ab initio molecular simulations with numeric atom-centered
  orbitals. \emph{Comput. Phys. Commun.} \textbf{2009}, \emph{180},
  2175--2196\relax
\mciteBstWouldAddEndPuncttrue
\mciteSetBstMidEndSepPunct{\mcitedefaultmidpunct}
{\mcitedefaultendpunct}{\mcitedefaultseppunct}\relax
\EndOfBibitem
\bibitem[Vollmer \latin{et~al.}(2001)Vollmer, Witte, and
  W{\"o}ll]{vollmer2001determination}
Vollmer,~S.; Witte,~G.; W{\"o}ll,~C. Determination of site specific adsorption
  energies of CO on copper. \emph{Catalysis letters} \textbf{2001}, \emph{77},
  97--101\relax
\mciteBstWouldAddEndPuncttrue
\mciteSetBstMidEndSepPunct{\mcitedefaultmidpunct}
{\mcitedefaultendpunct}{\mcitedefaultseppunct}\relax
\EndOfBibitem
\bibitem[Ren \latin{et~al.}(2009)Ren, Rinke, and Scheffler]{ren2009exploring}
Ren,~X.; Rinke,~P.; Scheffler,~M. Exploring the random phase approximation:
  Application to CO adsorbed on Cu(111). \emph{Phys. Rev. B} \textbf{2009},
  \emph{80}, 045402\relax
\mciteBstWouldAddEndPuncttrue
\mciteSetBstMidEndSepPunct{\mcitedefaultmidpunct}
{\mcitedefaultendpunct}{\mcitedefaultseppunct}\relax
\EndOfBibitem
\bibitem[Stroppa and Kresse(2008)Stroppa, and Kresse]{stroppa2008shortcomings}
Stroppa,~A.; Kresse,~G. The shortcomings of semi-local and hybrid functionals:
  what we can learn from surface science studies. \emph{New J. Phys.}
  \textbf{2008}, \emph{10}, 063020\relax
\mciteBstWouldAddEndPuncttrue
\mciteSetBstMidEndSepPunct{\mcitedefaultmidpunct}
{\mcitedefaultendpunct}{\mcitedefaultseppunct}\relax
\EndOfBibitem
\bibitem[Paier \latin{et~al.}(2007)Paier, Marsman, and Kresse]{paier2007does}
Paier,~J.; Marsman,~M.; Kresse,~G. Why does the B3LYP hybrid functional fail
  for metals? \emph{J. Chem. Phys.} \textbf{2007}, \emph{127}\relax
\mciteBstWouldAddEndPuncttrue
\mciteSetBstMidEndSepPunct{\mcitedefaultmidpunct}
{\mcitedefaultendpunct}{\mcitedefaultseppunct}\relax
\EndOfBibitem
\bibitem[Janesko \latin{et~al.}(2009)Janesko, Henderson, and
  Scuseria]{janesko2009screened}
Janesko,~B.~G.; Henderson,~T.~M.; Scuseria,~G.~E. Screened hybrid density
  functionals for solid-state chemistry and physics. \emph{Phys. Chem. Chem.
  Phys.} \textbf{2009}, \emph{11}, 443--454\relax
\mciteBstWouldAddEndPuncttrue
\mciteSetBstMidEndSepPunct{\mcitedefaultmidpunct}
{\mcitedefaultendpunct}{\mcitedefaultseppunct}\relax
\EndOfBibitem
\bibitem[Stoodley \latin{et~al.}(2024)Stoodley, Rochford, Lee, Klein, Duncan,
  and Maurer]{stoodley2024structure}
Stoodley,~M.~A.; Rochford,~L.~A.; Lee,~T.-L.; Klein,~B.~P.; Duncan,~D.~A.;
  Maurer,~R.~J. Structure of Graphene Grown on Cu (111): X-Ray Standing Wave
  Measurement and Density Functional Theory Prediction. \emph{Phys. Rev. Lett.}
  \textbf{2024}, \emph{132}, 196201\relax
\mciteBstWouldAddEndPuncttrue
\mciteSetBstMidEndSepPunct{\mcitedefaultmidpunct}
{\mcitedefaultendpunct}{\mcitedefaultseppunct}\relax
\EndOfBibitem
\bibitem[Perdew \latin{et~al.}(1996)Perdew, Burke, and
  Ernzerhof]{perdew1996generalized}
Perdew,~J.~P.; Burke,~K.; Ernzerhof,~M. Generalized gradient approximation made
  simple. \emph{Phys. Rev. Lett.} \textbf{1996}, \emph{77}, 3865\relax
\mciteBstWouldAddEndPuncttrue
\mciteSetBstMidEndSepPunct{\mcitedefaultmidpunct}
{\mcitedefaultendpunct}{\mcitedefaultseppunct}\relax
\EndOfBibitem
\bibitem[Larsen \latin{et~al.}(2017)Larsen, Mortensen, Blomqvist, Castelli,
  Christensen, Du{\l}ak, Friis, Groves, Hammer, Hargus, \latin{et~al.}
  others]{larsen2017atomic}
Larsen,~A.~H.; Mortensen,~J.~J.; Blomqvist,~J.; Castelli,~I.~E.;
  Christensen,~R.; Du{\l}ak,~M.; Friis,~J.; Groves,~M.~N.; Hammer,~B.;
  Hargus,~C.; others The atomic simulation environment—a Python library for
  working with atoms. \emph{J. Condens. Matter Phys.} \textbf{2017}, \emph{29},
  273002\relax
\mciteBstWouldAddEndPuncttrue
\mciteSetBstMidEndSepPunct{\mcitedefaultmidpunct}
{\mcitedefaultendpunct}{\mcitedefaultseppunct}\relax
\EndOfBibitem
\bibitem[Head-Gordon and Tully(1995)Head-Gordon, and Tully]{head1995molecular}
Head-Gordon,~M.; Tully,~J.~C. Molecular dynamics with electronic frictions.
  \emph{J. Chem. Phys.} \textbf{1995}, \emph{103}, 10137--10145\relax
\mciteBstWouldAddEndPuncttrue
\mciteSetBstMidEndSepPunct{\mcitedefaultmidpunct}
{\mcitedefaultendpunct}{\mcitedefaultseppunct}\relax
\EndOfBibitem
\bibitem[Kubo \latin{et~al.}(1991)Kubo, Toda, and
  Hashitsume]{kubo1991statistical}
Kubo,~R.; Toda,~M.; Hashitsume,~N. \emph{Statistical Physics II}, 2nd ed.;
  Springer Series in Solid-State Sciences; Springer Berlin, Heidelberg, 1991;
  Vol.~31; pp XVI, 279, Original Japanese edition published by Iwanami Shoten,
  Publishers, Tokyo, 1978\relax
\mciteBstWouldAddEndPuncttrue
\mciteSetBstMidEndSepPunct{\mcitedefaultmidpunct}
{\mcitedefaultendpunct}{\mcitedefaultseppunct}\relax
\EndOfBibitem
\end{mcitethebibliography}

\end{document}